\documentclass{aa} 

\usepackage{color}
\usepackage{graphicx}
\usepackage{natbib,twoopt}
\usepackage[varg]{txfonts}
\usepackage{gensymb}
\usepackage{lscape}

\bibpunct{(}{)}{;}{a}{}{,}

\renewcommand\labelenumi{(\theenumi)}

\def\hi{\relax \ifmmode {\mbox H\,\texjtsc{i}}\else H\,{\scshape i}\fi}
\def\hii{\relax \ifmmode {\mbox H\,\textsc{ii}}\else H\,{\scshape ii}\fi}
\def\nii{\relax \ifmmode {\mbox N\,\textsc{ii}}\else N\,{\scshape ii}\fi}
\def\oi{\relax \ifmmode {\mbox O\,\textsc{i}}\else O\,{\scshape i}\fi}
\def\oii{\relax \ifmmode {\mbox O\,\textsc{ii}}\else O\,{\scshape ii}\fi}
\def\oiii{\relax \ifmmode {\mbox O\,\textsc{iii}}\else O\,{\scshape iii}\fi}
\def\cii{\relax \ifmmode {\mbox C\,\textsc{ii}}\else C\,{\scshape ii}\fi}
\def\ciii{\relax \ifmmode {\mbox C\,\textsc{iii}}\else C\,{\scshape iii}\fi}
\def\civ{\relax \ifmmode {\mbox C\,\textsc{iv}}\else C\,{\scshape iv}\fi}
\def\hei{\relax \ifmmode {\mbox He\,\textsc{i}}\else He\,{\scshape i}\fi}
\def\heii{\relax \ifmmode {\mbox He\,\textsc{ii}}\else He\,{\scshape ii}\fi}
\def\mgii{\relax \ifmmode {\mbox Mg\,\textsc{ii}}\else Mg\,{\scshape ii}\fi}
\def\sii{\relax \ifmmode {\mbox S\,\textsc{ii}}\else S\,{\scshape ii}\fi}
\def\neiii{\relax \ifmmode {\mbox Ne\,\textsc{iii}}\else Ne\,{\scshape iii}\fi}
\def\ariv{\relax \ifmmode {\mbox Ar\,\textsc{iv}}\else Ar\,{\scshape iv}\fi}
\def\ni{\relax \ifmmode {\mbox N\,\textsc{i}}\else N\,{\scshape i}\fi}
\def\ariii{\relax \ifmmode {\mbox Ar\,\textsc{iii}}\else Ar\,{\scshape iii}\fi}
\def\caii{\relax \ifmmode {\mbox Ca\,\textsc{ii}}\else Ca\,{\scshape ii}\fi}

\begin{document}


   \title{The shape of oxygen abundance profiles explored with MUSE: evidence for widespread deviations from single gradients}
   \titlerunning{Shape of oxygen abundance profiles explored with MUSE}
   
   \author{L.~S\'anchez-Menguiano\inst{1,2}\and S.~F.~S\'anchez\inst{3}\and I.~P\'erez\inst{2,4}\and T.~Ruiz-Lara\inst{5,6}\and L.~Galbany\inst{7}\and J.~P.~Anderson\inst{8}\and T.~Kr\"uhler\inst{9} \and H.~Kuncarayakti\inst{10,11} \and J.~D.~Lyman\inst{12}}
   \authorrunning{L.~S\'anchez-Menguiano et al.}

   \institute{Instituto de Astrof\'isica de Andaluc\'ia (CSIC), Glorieta de la Astronom\'ia s/n, Aptdo. 3004, E-18080 Granada, Spain\\
              \email{lsanchez@iaa.es}
         \and Dpto. de F\'isica Te\'orica y del Cosmos, Universidad de Granada, Facultad de Ciencias (Edificio Mecenas), E-18071 Granada, Spain
         \and Instituto de Astronom\'ia, Universidad Nacional Aut\'onoma de M\'exico, A.P. 70-264, 04510, M\'exico, D.F.
         \and Instituto Carlos I de F\'isica Te\'orica y computacional, Universidad de Granada, E-18071 Granada, Spain
         \and Instituto de Astrof\'isica de Canarias, Calle V\'ia L\'actea s/n, E-38205 La Laguna, Tenerife, Spain
         \and Universidad de La Laguna, Dpto. Astrof\'isica, E-38206 La Laguna, Tenerife, Spain
         \and PITT PACC, Department of Physics and Astronomy, University of Pittsburgh, Pittsburgh, PA 15260, USA 
         \and European Southern Observatory, Alonso de C\'ordova 3107, Casilla 19001, Santiago, Chile
         \and Max-Planck-Institut f\"{u}r extraterrestrische Physik, Giessenbachstra\ss e, 85748 Garching, Germany
         \and Finnish Centre for Astronomy with ESO (FINCA), University of Turku, Vaisalantie 20, 21500 Piikkio, Finland
         \and Tuorla Observatory, Department of Physics and Astronomy, University of Turku, Vaisalantie 20, 21500 Piikkio, Finland
         \and Department of Physics, University of Warwick, Coventry CV4 7AL, UK\\
             } 

   \date{Received 1 July 2017 / Accepted 2 October 2017}


\abstract{We characterised the oxygen abundance radial distribution of a sample of 102 spiral galaxies observed with VLT/MUSE using the O3N2 calibrator. The high spatial resolution of the data allowed us to detect 14345 \hii\,regions with the same image quality as with photometric data, avoiding any dilution effect. We developed a new methodology to automatically fit the abundance radial profiles, finding that 55 galaxies of the sample exhibit a single negative gradient. The remaining 47 galaxies also display, as well as this negative trend, either an inner drop in the abundances (21), an outer flattening (10), or both (16), which suggests that these features are a common property of disc galaxies. The presence and depth of the inner drop depends on the stellar mass of the galaxies with the most massive systems presenting the deepest abundance drops, while there is no such dependence in the case of the outer flattening. We find that the inner drop appears always around $\rm 0.5\,r_e$, while the position of the outer flattening varies over a wide range of galactocentric distances. Regarding the main negative gradient, we find a characteristic slope in the sample of $\alpha_{O/H} = -\,0.10\pm0.03\,\rm{dex}/r_e$. This slope is independent of the presence of bars and the density of the environment. However, when inner drops or outer flattenings are detected, slightly steeper gradients are observed. This suggests that radial motions might play an important role in shaping the abundance profiles. We define a new normalisation scale (`the abundance scale length', $r_{O/H}$) for the radial profiles based on the characteristic abundance gradient, with which all the galaxies show a similar position for the inner drop ($\sim0.5\,r_{O/H}$) and the outer flattening ($\sim1.5\,r_{O/H}$). Finally, we find no significant dependence of the dispersion around the negative gradient with any property of the galaxies, with values compatible with the uncertainties associated with the derivation of the abundances.} 


\maketitle

\section{Introduction}\label{sec:intro}

The study of the gas-phase chemical composition of spiral galaxies has proven to be an important tool to improve our knowledge of the evolution of these complex systems. Models for the formation of spiral galaxies agree on a scenario based on an inside-out growth of the discs as a result of the increased timescales of the gas infall with radius and the consequent radial dependence of the star formation rate \citep{matteucci1989, boissier1999}. In this context, the analysis of the spatial distribution of chemical abundances provides strong constraints for the main parameters describing these chemical evolution models \citep{koeppen1994, edmunds1995, tsujimoto1995, molla1996, molla1997, prantzos2000, chiappini2001, molla2005, fu2009, pilkington2012}.

Consistent with this inside-out paradigm, observational studies on the gas metallicity from \hii\,regions have found the presence of negative radial gradients across the discs of nearby galaxies \citep[][among many others]{searle1971, comte1975, smith1975, martin1992, vanzee1998}. It was only found decades later that this radial distribution may present some deviations from this simple negative gradient, namely a decrease of the abundance in the innermost regions and a flattening of the gradient in the outer parts \citep[e.g.][]{belley1992, martin1995, vilchez1996, roy1997}. Despite the wide variety of mechanisms proposed to explain the presence of these features (such as radial migration, \citealt{minchev2011,minchev2012}; or satellite accretion, \citealt{qu2011, bird2012}), their origins are still unclear.

This line of research significantly benefited from the development of integral field spectroscopic (IFS) techniques that allowed for better characterisation of the abundance distribution in galaxies thanks to the two-dimensional field-of-view (FoV) of these instruments. The first IFS studies that analysed statistically the properties of the \hii\,regions were presented by the PINGS project \citep{rosalesortega2010}, with the disadvantage of being limited to a dozen nearby galaxies. The use of these techniques on `survey mode', that offers the opportunity to carry out these studies on large samples of galaxies, was crucial to reach meaningful statistical results \citep{moustakas2010, rich2012, sanchez2012b, sanchez2014, ho2015, sanchezmenguiano2016, zinchenko2016, belfiore2017}. 

For instance, studies using data from IFS surveys have allowed us to find that spiral galaxies in the Local Universe present a characteristic abundance gradient when normalised to a physical scale of the discs \citep{sanchez2012b,sanchez2014}. \citet{ho2015}, based on the use of simple chemical evolution models, proposed that this common gradient was the result of common gas and stellar surface density profiles under the co-evolution of gas, stars, and metals during the mass growth of disc galaxies. IFS surveys have also allowed us to find a relation between the inner abundance drop presented in spiral galaxies and the galaxy mass, being this feature more common in more massive galaxies \citep{sanchezmenguiano2016}. Regarding the study of azimuthal abundance variations in spiral galaxies, \citet{zinchenko2016} suggested that there is no significant global azimuthal asymmetry in the oxygen abundance distribution, establishing a limit of $\sim0.05$ dex for these asymmetries, although works analysing in detail individual galaxies indicate otherwise \citep{sanchezmenguiano2016b, vogt2017}.

However, despite the great advantages and scientific legacy that these instruments offer, they also present some limitations. Most of them do not have the sufficient spatial resolution needed to resolve small morphological structures of the galaxies such as the individual \hii\,regions. \citet{mast2014} showed that the loss of spatial resolution reduces significantly the number of detections of these star-forming regions. This fact hampers a good characterisation of the distribution of their properties across the galaxy extent and, according to the authors, can even lead to erroneous results. 

The advent of new instruments with high spatial resolution like GMOS \citep[Gemini Multiobject Spectrograph,][]{allington2002} or OASIS\footnote{http://cral.univ-lyon1.fr/labo/oasis/present/} (Optically Adaptive System for Imaging Spectroscopy) has allowed us to overcome these limitations in the Local Universe. However, the limited FoV of these instruments reduces the collection of data to particular regions of nearby galaxies or forces us to move to high redshift to observe the entire extent of galaxies, to the detriment of the effective physical spatial resolution. 

In this regard, the development of MUSE \citep[Multi Unit Spectroscopic Explorer,][]{bacon2010} has meant a new revolution in the progress of IFS by combining a large FoV with high spatial resolution, helping us to go further in the study of the gas-phase spatially resolved chemical distribution. These two characteristics combined allow us to increase the number of \hii\,regions detected in individual galaxies with respect to previous studies and better trace their distribution across the galaxy discs. Moreover, thanks to the high spatial resolution, unprecedented in two-dimensional spectroscopic studies, we can avoid any dilution effect in the detection of the \hii\,regions and segregate them with the same quality as with photometric data. This way, we can carry out a better characterisation of the oxygen abundance radial distribution, which is the purpose of this work.

The capabilities of MUSE data for these types of studies were already demonstrated in a pilot project analysing the radial abundance gradient of NGC~6754 \citep{sanchez2015a}. There, the authors constructed one of the largest catalogue of \hii\,regions with available spectroscopic information in a single galaxy up to date, comprising 396 individual ionised sources, and used it to derive the radial abundance gradient of the galaxy. Since then, new works on the gas-phase abundance distribution based on MUSE data have arisen \citep[e.g.][]{kreckel2016,sanchezmenguiano2016b,vogt2017}, confirming the potential of this instrument for such studies.

In this work, we aim to extend the characterisation of the oxygen abundance radial profile to a large sample of spiral galaxies. For this purpose we develop a new methodology to automatically fit these profiles taking into account the different shapes observed. This procedure allows us to detect the presence of any possible deviation in the radial distribution with respect to the universal negative gradient (i.e. the inner drop and the outer flattening). Thanks to the sample size, by analysing the characteristics of the derived profiles for all the galaxies we are able to state meaningful statistical conclusions. Moreover, the study of possible trends with different properties of the galaxies allows us to shed light into the origin of these deviations, and the role of such properties in the chemical evolution of the galaxies.

The structure of the paper is organised as follows: Section~\ref{sec:sample} provides a description of the properties of the sample as well as the data used in this study. In Sect.~\ref{sec:analysis} we explain the analysis required to detect the \hii\,regions and derive the corresponding oxygen abundance radial distribution. The results of the analysis are shown in Sect.~\ref{sec:results}. Along this section, we describe the shape of the abundance profiles (Sect.~\ref{sec:shape}) and study the dependence of the slope of the gradient on different properties of the galaxies (Sect.~\ref{sec:slope}). In addition, we analyse the location of the inner abundance drop and the outer flattening when existing (Sect.~\ref{sec:dropflattening}), we explore the existence of a common oxygen abundance gradient (Sect.~\ref{common}), we define a new normalisation scale for the radial profiles based on this characteristic abundance gradient (Sect.~\ref{masses}), and we study the dispersion relative to the abundance gradient (Sect.~\ref{sec:scatter}). Finally, the main conclusions are given in Sect.~\ref{sec:conclusions}.


\section{Data and galaxy sample}\label{sec:sample}

In this work we used a sample of galaxies part of the AMUSING \citep[All-weather MUse Supernova Integral-field Nearby Galaxies, PI Anderson/Galbany;][]{galbany2016} survey (see affiliated programmes in the acknowledgements). The data were collected with the MUSE instrument \citep{bacon2010, bacon2014}, mounted on the Unit 4 telescope (UT4) at the Very Large Telescope (VLT) of the Cerro Paranal Observatory. In the Wide Field Mode, this integral-field spectrograph presents a FoV of approximately $1'\times1'$ and a pixel size of $0.2''$, which limits the spatial resolution of the data to the atmospheric seeing during the observations. Regarding the covered wavelength range, MUSE spans from $4750\AA$ to $9300\AA$, with a spectral sampling of $1.25\AA$ and a spectral resolution between $1800-3600$ (from the blue edge of the spectrum to the red one).

The median seeing during the observations of the analysed dataset is $1.1''$ (80\% of the values ranging between $0.8''$ and $1.8''$), corresponding to a median physical resolution of $\sim460$~pc. The typical exposure time of each pointing is $\sim50$ minutes and the limiting surface brightness in $g$-band covers the range between $22.6-24.9$ mag/arcsec$^2$ (10-th and 90-th percentiles), with a median value of $24.0$ mag/arcsec$^2$. Figure~\ref{fig:spectra} shows two examples of typical spectra of \hii\,regions, illustrating the quality of the data.

The data analysed here have been reduced using version 1.2.1 of the MUSE pipeline \citep{weilbacher2014} and the Reflex environment \citep{freudling2013}. This pipeline layout includes the standard processes like bias subtraction and flat-fielding, flux and wavelength calibration. The sky subtraction was performed using either an offset pointing to a blank sky position, or blank sky regions within the science frames, and making use of algorithms from the Zurich Atmosphere Purge package \citep[ZAP;][]{soto2016}. The effects of Galactic extinction were also corrected. Finally, a geometrical calibration was performed to distribute the individual slices of the image within the FoV and reconstruct the datacube. More information about the data reduction can be found in \citet{galbany2016} and \citet{kruhler2017}.

For some galaxies, the observations were split into different pointings covering different areas of the galaxies to construct a mosaic. A set of $\sim100\,000$ spectra is delivered per pointing, reaching up to an impressive number of $\sim600\,000$ spectra for the most extended galaxies (mosaics composed by six pointings). 

\begin{figure}
\centering
\resizebox{0.95\hsize}{!}{\includegraphics{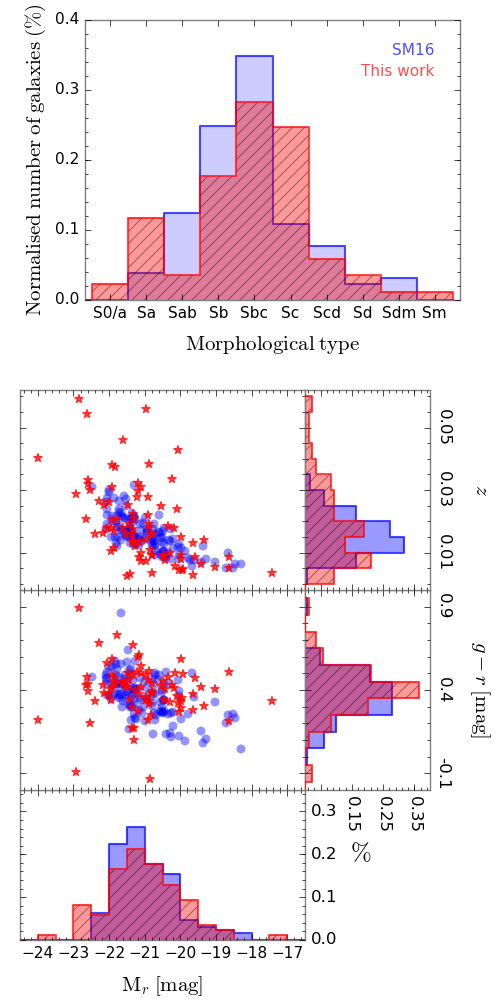}}
\caption{Characterisation of the analysed sample (red dashed histograms and stars) as compared with the one used in \citet[][blue solid histograms and circles]{sanchezmenguiano2016}. The plot shows the distribution of morphological types ({\it top}), and the redshift and $g-r$ colour distributions as a function of the absolute $r$-band magnitudes ({\it bottom}). The frequency histograms for these three parameters are also shown.}
\label{fig:hists_sample}
\end{figure}

\vspace{0.5cm}
AMUSING is an ongoing project aimed at studying the environments of supernovae (SNe) by means of the analysis of a large number of nearby SN host galaxies (redshifts typically between 0.005 and 0.06, with exceptional cases up to 0.1; see Fig.~\ref{fig:hists_sample} for the range covered by the subsample analysed in this work). The AMUSING sample comprises up to date more than 300 galaxies. 227 of these galaxies are part of the observing campaign (expanding four semesters so far) carried out by the project. The remaining ones have been included after a carefull search in the European Southern Observatory (ESO) archive fulfiling the criteria imposed by AMUSING (galaxies observed with MUSE where SNe have been discovered). As a consequence, the mother sample is composed by a wide variety of galaxy types with the only characteristic in common of hosting or having hosted a SN. Because of the large percentage of spiral galaxies in the sample we can study for the first time the oxygen abundance distribution of the gas-phase in a meaningful statistical sample with an unprecedented spatial resolution as the one provided by MUSE. In order to do that, we first removed galaxies with low image quality (visually presenting low signal in the source and/or with seeing values above $2''$) and those that are too small (projected semimajor axis in the sky smaller than $\sim15''$) to properly derive the abundance radial distribution without mixing information (very large physical size of the detected \hii\,regions). Galaxies for which the centre of the galactic discs is not mapped were also discarded from the sample. From the remaining galaxies we then selected the subset that fulfils the following criteria:
\begin{enumerate}
\item Spiral galaxies with morphological types between Sa and Sm, including barred galaxies.
\item Intermediate to low inclined galaxies ($i < 70 \degree$) to avoid uncertainties induced by inclination effects.
\item Galaxies with at least 10 \hii\,regions to properly derive the abundance radial profile (see Sect.\ref{sec:analysis}).
\item Galaxies with enough radial coverage of the abundance profiles (at least 0.5 disc effective radii, see Sect.\ref{sec:analysis3}).
\end{enumerate}
Regarding the morphological type criterion, the classification is based on the information provided in the extragalactic database {\it HyperLeda}\footnote{http://leda.univ-lyon1.fr} \citep{makarov2014}. However, due to the higher quality of the analysed data over the previously available ones, a visual inspection of the images reconstructed from the MUSE datacubes was done to check the galaxies. Indeed, two galaxies that were classified by {\it HyperLeda} as elliptical or lenticular were kept in the sample because of the presence of a clear spiral structure observed with our data (PGC~55442 and NGC~232), and two galaxies classified as spirals were discarded for actually being lenticular (NGC~4772 and NGC~6962). Moreover, for the galaxies with no morphological information in the mentioned database (18), a visual selection was performed, all of them being finally included. After imposing these restrictions, the sample is reduced to 102 galaxies. General information about these galaxies including the morphological type, redshift, stellar mass and disc effective radius is given in Table~\ref{table1} (Appendix~\ref{sec:appendix1}). Furthermore, in Appendix~\ref{sec:appendix4} we provide a graphic characterisation of the sample showing two RGB colour images of the galaxies (using SDSS $r$-, $i$- and $z$-band images recovered from the data, and narrow-band images centred in the emission lines \mbox{[\nii]~$\lambda6584$}, H$\alpha$, and \mbox{[\oiii]~$\lambda5007$}).

While the AMUSING mother sample has not been defined based on their galaxy properties, here we argue that the sample used in this work is representative of the galaxy population in the Local Universe. A criterion based on the occurrence on the galaxy being host to a SN is proposed as a valid selection to find galaxies properly populating the galaxy mass distribution (galaxies hosting SNe~Ia) and the typical observed ranges of star formation rates (galaxies hosting core-collapse SNe). In Fig.~\ref{fig:hists_sample} we demonstrate this fact by comparing our sample of galaxies with a well-defined sample presented in a previous work of the oxygen abundance radial distribution \citep[][hereafter SM16]{sanchezmenguiano2016} based on the CALIFA survey \citep[Calar Alto Legacy Integral Field Area Survey,][]{sanchez2012a}. We should note that the comparison sample was extracted from a well defined, statistically significant sample and representative of galaxies in the Local Universe \citep{walcher2014}. The top panel of Fig.~\ref{fig:hists_sample} shows the normalised distribution of morphological types of the 84 galaxies of the sample for which {\it HyperLeda} information on this regards is available (red dashed histogram) and the galaxies used in SM16 (blue solid histogram). The sample is dominated by galaxies of intermediate morphological types ranging from Sb to Sc, although spirals of all morphological types are found. Despite the lack of Sab galaxies in favour of Sa systems with respect to the SM16 sample, the coverage of morphological types is very similar in both samples. The bottom panel of Fig.~\ref{fig:hists_sample} shows the redshift and $g-r$ colour distributions as a function of the absolute $r$-band magnitudes, together with their corresponding frequency histograms for the current analysed sample (red stars and red dashed histograms) and the SM16 sample (blue circles and blue solid histograms). The absolute $r$-band magnitudes and $g-r$ colours are computed from the flux density in $r$-band and $g$-band images recovered from the data. The coverage of both samples is again very similar for all the parameters, with the exception of the tail towards higher redshifts presented in our sample, as no redshift criterion was applied to define our sample. If we do not consider the galaxies with $z > 0.03$, restricting ourself to the range covered by the SM16 sample, a Kolmogorov-Smirnov test (KS-test) yields p-values above the significance level ($5\%$) for the three parameters, concluding that the two samples present  statistically similar distributions. 

Therefore, we conclude that the sample analysed here is suitable for carrying out a detailed study of the oxygen abundance radial distribution in spiral galaxies, and for drawing statistical conclusions applicable to spiral galaxies in the Local Universe.


\section{Analysis}\label{sec:analysis}

In this section we describe the process followed to derive the oxygen abundance radial distribution for each galaxy in the sample. This procedure includes the detection of the \hii\,regions, the measurement of the emission line fluxes used to determine the oxygen abundances, and the derivation of the radial distribution of these abundances.

\subsection{Detection of ionised regions}\label{sec:analysis1}

We detected the `candidates' for \hii\,regions (clumpy ionised regions in general) and extracted their corresponding spectra using \mbox{\scshape HIIexplorer}\footnote{\url{http://www.astroscu.unam.mx/~sfsanchez/HII\_explorer}}. A detailed description of the implementation of this algorithm on CALIFA data \citep{sanchez2012a} can be found in \citet{sanchez2012b}. However, some modifications in the code were done to adapt the algorithm to high-resolution data. A few of these adjustments were already presented in \citet{sanchez2015a}, but the fully updated version of \mbox{\scshape HIIexplorer} compatible with MUSE data is described here. 

This tool assumes that \hii\,regions are peaky and isolated structures where the ionised gas emission, particularly H$\alpha$, is strong enough as to exceed the stellar continuum emission and the diffuse ionised gas emission across the galaxy. On the other hand, \mbox{\scshape HIIexplorer} also considers extragalactic \hii\,regions to have a typical physical size of about a few hundred parsecs \citep{gonzalezdelgado1997, oey2003, lopez2011}. Taking into account these assumptions, we can summarise the main steps of the process as follows: 
\begin{enumerate}
\item First we create an H$\alpha$ narrow-band image of each galaxy, using a width of 120 \AA\, with the central wavelength shifted to the redshift of the galaxy, that is given as an input to \mbox{\scshape HIIexplorer}.
\item Using this image, the algorithm detects the brightest pixel in the map and then adds all the adjacent pixels up to a distance of $0.9''$ if their fluxes exceed 1\% of the peak intensity. After the first region is detected, its area is masked from the input image and the procedure is repeated until no peak with a flux exceeding the median H$\alpha$ emission flux of the galaxy is found. If two peaks are close to each other ($d < 1.8''$), the algorigthm will assign to the region with the brightest peak pixels that might belong to the faintest one. In order to reassociate pixels within this `overlapping' region, \mbox{\scshape HIIexplorer} computes the flux ratio between those pixels and both peaks. These pixels will belong to the region presenting the largest ratio providing such value is above 10\%. This criterion avoids the extension of regions into others, producing a `straight' division between them. The final result is a segmentation map describing the pixels associated with each detected ionised region. 
\item Finally, the integrated spectrum corresponding to each segmented region is extracted from the original datacube, and the corresponding position table of the detected areas is provided.
\end{enumerate}

\begin{figure*}
\centering
\resizebox{0.9\hsize}{!}{\includegraphics{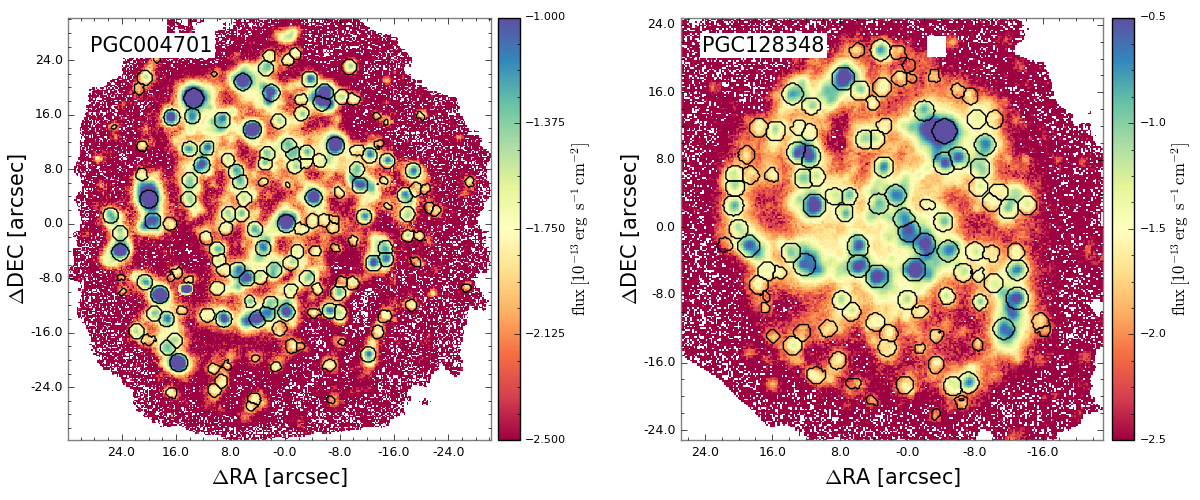}}
\caption{H$\alpha$ maps in units of (log$_{10}$) $\rm 10^{-13} \,erg \;s^{-1}\,cm^{-2}$ derived for two galaxies of the sample, PGC004701 (left) and PGC128348 (right), together with the ionised regions detected by \mbox{\scshape HIIexplorer} shown as black segmented contours.}
\label{fig:seg}
\end{figure*}

For this work, only the regions with $\rm S/N>3$ in H$\alpha$ emission were considered to guarantee the detection of the \hii\,regions. Figure~\ref{fig:seg} displays examples of the H$\alpha$ intensity maps of two galaxies of the sample, PGC004701 (left) and PGC128348 (right). In these images, the ionised regions detected by \mbox{\scshape HIIexplorer} are superimposed as black segmented contours. In these particular galaxies, we find 157 and 113 ionised regions, respectively. On average, a total of 140 regions are detected for each galaxy of our sample, but there are cases where the number of detections increases up to several hundred. The same H$\alpha$ intensity maps for the rest of the sample are shown in Appendix~\ref{sec:appendix4}. In Fig.~\ref{fig:num_regions} we show the distribution of the number of ionised regions detected per galaxy. The wide variety of values found reflects the diversity of galaxies present in the sample, from very large face-on spiral galaxies to smaller and more inclined systems. However, for all of them the sufficiently high number of detected \hii\,regions guarantees a proper characterisation of the radial oxygen abundance gradient. Finally, the total number of regions we have detected in the sample is 14345.

\begin{figure}
\centering
\resizebox{\hsize}{!}{\includegraphics{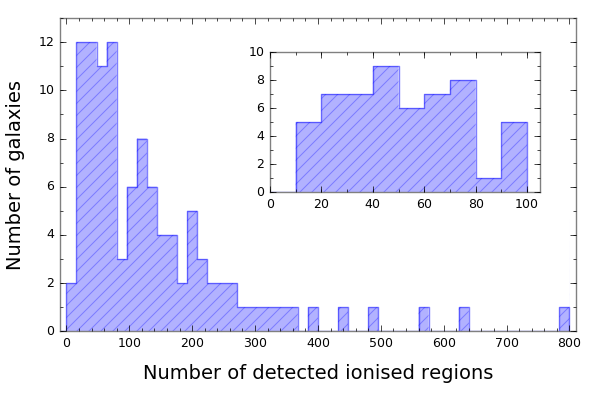}}
\caption{Distribution of the number of ionised regions detected by \mbox{\scshape HIIexplorer} for each galaxy. The inset represents the same distribution for the galaxies with less than 100 detected regions.}
\label{fig:num_regions}
\end{figure}

\subsection{Determination of the gas abundance values}\label{sec:analysis2}

Prior to the determination of the emission line fluxes needed to measure gas-phase abundances, the stellar contribution must be estimated and subtracted from the spectrum to derive a pure gas spectrum for each individual region detected by \mbox{\scshape HIIexplorer}. 

\subsubsection{Measurement of the emission lines}\label{sec:analysis2a}

Several tools have been developed to model the underlying stellar population and decouple it from the gaseous emission lines \citep[e.g.][]{cappellari2004, cidfernandes2005, ocvirk2006a, ocvirk2006b, sarzi2006, koleva2009, sanchez2011}. Most of them are based on the assumption that the star formation history (SFH) of a galaxy can be approximated as a sum of discrete star formation bursts and, therefore, that the stellar spectrum can be considered as the result of the combination of spectra of different simple stellar populations (SSP) with different ages and metallicities.

In this work, we make use of a fitting package named FIT3D\footnote{\url{http://www.astroscu.unam.mx/~sfsanchez/FIT3D}}. FIT3D is now part of a more complete spectroscopic analysis pipeline named {\scshape Pipe3D} \citep{sanchez2016a}, developed to characterise the properties of both the stellar populations and the ionised gas. In the modelling of the continuum emission, FIT3D uses an SSP template grid combining the GRANADA models from \citet{gonzalezdelgado2005} for $t < 63$ Myr with those provided by the MILES project \citep{sanchezblazquez2006,vazdekis2010,falconbarroso2011} for older ages \citep[following][]{cidfernandes2013}. This grid comprises a total of 156 individual populations covering 39 stellar ages between 0.001 and 14.1 Gyr and four metallicities between 0.004 and 0.03. In this manner, FIT3D fits each spectrum by a linear combination of the SSP templates after correcting for the appropriate systemic velocity and velocity dispersion and taking into account the effects of dust attenuation. To correct for the stellar dust extinction we adopted the \citet{cardelli1989} law with $R_{V} = 3.1$. 

\begin{figure*}
\begin{center}
\resizebox{\hsize}{!}{\includegraphics{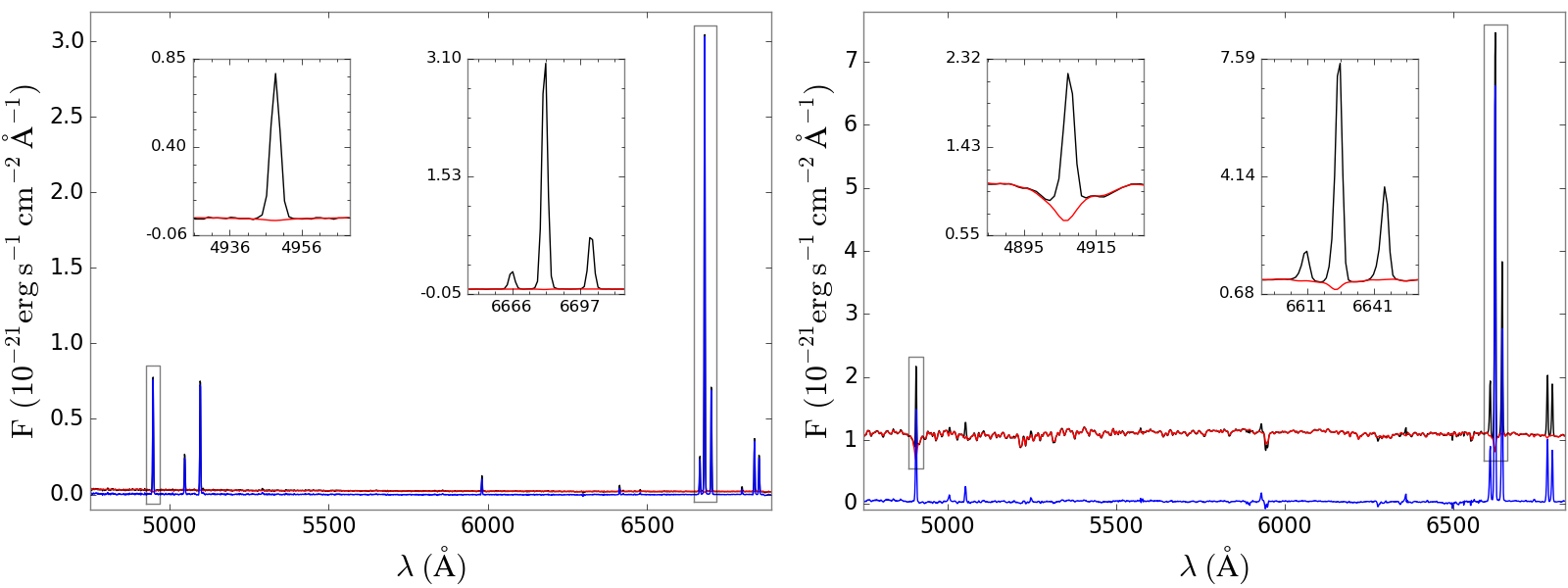}}
\caption{Two typical spectra of \hii\,regions observed in the sample with different levels of stellar contribution. The black line represents the observed spectrum, along with the best fitted stellar population performed by FIT3D (SSP model, red line). The pure emission line spectrum after subtracting the SSP model is shown as a blue line. The insets are focussed on the $H\beta$ and $H\alpha-NII$ spectral regions to highlight the quality of both the spectra and the subtraction of the underlying stellar continuum.}
\label{fig:spectra}
\end{center}
\end{figure*}

After the stellar component is subtracted, FIT3D measures the emission line fluxes by performing a multi-component fitting using a single Gaussian function per emission line plus a low-order polynomial function. When more than one emission line was fitted simultaneously (e.g. for doublets and triplets like the [\nii] lines), the systemic velocity and velocity dispersion were forced to be equal to decrease the number of free parameters and increase the accuracy of the deblending process. The measured line fluxes include all lines required to determine the gas metallicity using strong-line methods that fall within the wavelength range covered by MUSE, that is, H$\alpha$, H$\beta$, \mbox{[\oiii]~$\lambda4959$}, \mbox{[\oiii]~$\lambda5007$}, \mbox{[\nii]~$\lambda6548$}, \mbox{[\nii]~$\lambda6584$}, \mbox{[\sii]~$\lambda6717,$} and \mbox{[\sii]~$\lambda6731$}. FIT3D provides the intensity, equivalent width (EW), systemic velocity, and velocity dispersion for each emission line. 

The estimation of the errors of the parameters is done by performing a Monte Carlo (MC) simulation, where the original emission line spectrum is perturbed by a noise spectrum that includes both the original estimated errors and the uncertainties in the best fit SSP model. The uncertainties of the subtraction of the underlying stellar population are therefore propagated to the parameters derived for the emission lines. 

Figure~\ref{fig:spectra} shows two typical observed spectra from the \hii\,regions analysed in the sample (black line), along with the best fitted stellar population (SSP model, red line) and the pure emission line spectrum after subtracting the SSP model (blue line). The insets are focussed on the $H\beta$ and $H\alpha-NII$ spectral regions to highlight the quality of both the spectra and the subtraction of the underlying stellar continuum performed by FIT3D. These examples show different levels of contribution of the stellar continuum in the spectrum of the \hii\,regions. In cases where the stellar contribution is low, its effect on the emission lines recovery is negligible (compare blue and black lines of the left panel). On the other hand, if the contribution is significant (see right panel), the performance of FIT3D is quite remarkable, being able to properly model the underlying stellar absorption in features such as H$\alpha$ and H$\beta$.

The entire procedure of fitting and subtracting the underlying stellar population and measuring the emission lines using FIT3D, together with other algorithms of {\scshape Pipe3D} are extensively described in \citet{sanchez2011}, \citet{sanchez2016a} and \citet{sanchez2016b}. 

\subsubsection{Selection of the \hii\,regions}\label{sec:analysis2b}

{\scshape HIIexplorer} allows us to detect clumpy ionised regions. However, not all these regions may be associated with star formation as some of them may be ionised by other sources as active galactic nuclei (AGNs) or compact shocks. The intensity of strong lines has been broadly used to differentiate between different types of emission according to their main excitation source (i.e. starburst or AGN) throughout the so-called diagnostic diagrams. The most common, and that used here, was proposed by \citet[][hereafter BPT diagram]{baldwin1981}. This diagram makes use of the \mbox{[\nii]~$\lambda6584$/H$\alpha$} and \mbox{[\oiii]~$\lambda5007$/H$\beta$} line ratios, which are less affected by dust attenuation because of their proximity in wavelength space. For this diagram, different demarcation lines have been proposed to distinguish between \hii\,regions and AGNs. The most popular ones are the \citet{kewley2001} and \citet{kauffmann2003} curves. The former line was derived theoretically using photoionisation models and the latter one is based on the analysis of the integrated spectra of SDSS galaxies. However, it has been found that the \citet{kauffmann2003} curve excludes certain kinds of SF regions already observed in galaxies (\citealt{kennicutt1989b,ho1997} and, more recently, \citealt{perezmontero2009} and \citealt{sanchez2014}). For this reason, and in order to avoid biasing our sample towards classical disc regions, we adopted the \citet{kewley2001} curve to select our \hii\,regions.

Following the procedure proposed in \citet{sanchez2014}, in addition to the use of the BPT diagram we also adopted the WHAN diagram \citep[W$_{\rm H\alpha}$ versus \mbox{[\nii]/H$\alpha$},][]{cidfernandes2011} to guarantee the exclusion of other sources of ionisation that might populate the SF area, such as weak AGNs, shocks, and/or post-AGBs stars. This diagram uses the EW(H$\alpha$) to take into account weak AGNs and `retired' galaxies, that is, galaxies that have stopped forming stars and are ionised by hot low-mass evolved stars. We have been more restrictive in the EW range than \citet{cidfernandes2011} and established the limit in 6~\AA\, to assure a significant percentage (20\%) of young stars contributing to the emission of the \hii\,regions \citep[given the strong correlation between both parameters, see][]{sanchez2014}.

\subsubsection{Measurement of the oxygen abundances}\label{sec:analysis2c}

The most direct and reliable method developed to measure abundances from observed spectra requires using temperature-sensitive line ratios such as \mbox{[\oiii]~$\lambda\lambda4959,5007/[\oiii]~\lambda4363$}. This is known as the $T_e-$method \citep{peimbert1969,stasinska1978,pagel1992,vilchez1996,izotov2006}. However, the weakness of some of these lines (that become even fainter as the metallicity increases) hampers their detection and makes this method only applicable to very nearby galaxies for which very high S/N spectra are attainable.

For this reason, alternative procedures based on relations between metallicity and the intensity of strong and more easily observable lines have been proposed. Thus, two families of calibrators have been developed and are widely used today: empirical calibrators based on direct estimations of oxygen abundances \citep{zaritsky1994, pilyugin2000, pettini2004, perezmontero2005, pilyugin2005, pilyugin2010, marino2013} and calibrators based on photoionisation models \citep{mcgaugh1991, kewley2002, kobulnicky2004, dopita2006, dopita2013, perezmontero2014, dopita2016}. It is important to note that $\rm T_e$-based empirical calibrations provide results that are at least $0.2-0.4$~dex lower than strong-line methods based on photoionisation models (see \citealt{lopezsanchez2010, lopezsanchez2012} for extended discussion of these issues).

In this work we adopted the empirical calibrator based on the O3N2 index that was first introduced by \citet{alloin1979} and afterwards modified by \citet{pettini2004}:

\begin{equation}
{\rm O3N2} =  \log\left(\frac{[\oiii] \lambda5007}{{\rm H}\beta} \times \frac{{\rm H}\alpha}{[\nii] \lambda6584}\right)
.\end{equation}

The close distance in wavelength between the lines of both ratios makes this index barely affected by dust attenuation. In addition, it presents a monotonic dependence on the abundance and uses emission lines covered by the MUSE wavelength range (other lines such as \mbox{[\oii]~$\lambda3727$} are out of its range). For this index, we used the calibration proposed by \citet[][hereafter M13]{marino2013}, where \mbox{$12+\log\left({\rm O/H}\right) = 8.533 - 0.214 \,\times\, {\rm O3N2}$}. This calibration uses $T_e$-based abundances of $\sim 600$ \hii\, regions from the literature together with new measurements from the CALIFA survey \citep{sanchez2012a}, providing the most update and accurate calibration to date for this index. The increase in the number of regions used in this calibration is especially significant in the high-metallicity regime, where previous calibrators based on this index lack of high-quality observations of \hii\,regions with auroral lines at this high-abundance end \citep[e.g.][]{pettini2004,perezmontero2009}.

We note that the linear regression derived in M13 between the O3N2 parameter and the oxygen abundance is valid for the interval $\rm 8.17 < 12+\log(O/H) < 8.77$. \hii\,regions presenting values outside this range were therefore not considered in the analysis. The derived abundances have a calibration error of $0.08$ dex, much larger than the typical one associated with the errors in the measured emission lines that are taken into account via MC simulations ($\sim0.02$ dex, being the errors in the involved emission lines between 1\% and 10\% the flux values).

For completeness, and in order to verify the reliability of our results and their dependence on the adopted method, we have also made use of two other calibrators. However, the limited wavelength range in the blue regime covered by MUSE does not allow us to use more well-known calibrators based on emission lines that are out of this range (in particular \mbox{[\oii]~$\lambda3727$}). Therefore, we have selected other less common abundance indicators, like the ones based on the sulphur lines. In particular, we make use of (i) the calibration described in \citet[][hereafter D16]{dopita2016}; and (ii) the one proposed in \citet[][]{marino2013} for the N2 index (hereafter M13-N2). The D16 calibrator is based on photoionisation models and makes use of H$\alpha$, \mbox{[\nii]~$\lambda6584$} and \mbox{[\sii]~$\lambda\lambda6717,6731$} emission lines. This calibrator is almost linear up to an abundance of $12+\log\left({\rm O/H}\right) = 9.05$ and it is independent of reddening (it invokes lines located close together in wavelength). Finally, the M13-N2 indicator is based in the N2 index, defined as $N2 = \log\left([\nii]\lambda6584/H\alpha\right)$. This calibrator is also independent of reddening and has a monotonic behaviour, but presents the disadvantage of saturating in the high-metallicity regime.

\subsection{Radial distribution of the oxygen abundances}\label{sec:analysis3}

To derive the radial distribution of abundances it is necessary to determine the deprojected galactocentric distances of the \hii\,regions to remove possible effects of inclination in the analysis. To do this, we need to obtain the two deprojection angles of the disc: the position angle (PA) and the inclination angle ($i$). 

These angles were derived by fitting ellipses of variable ellipticity and position angle to the outermost isophotes of the galaxies in a $g$-band image recovered from the data, by means of the {\tt ellipse IRAF} task. For the galaxies that are larger than the FoV of MUSE or very close to its limit (32) a proper characterisation of the outermost isophotes is highly hampered using these data and therefore, the Digitized Sky Survey (DSS) red-band (POSS-II F, \citealt{reid1991}) images were used instead. Table~~\ref{table1} (Appendix~\ref{sec:appendix1}) lists the values of the position and inclination angles for all galaxies in our sample. Following \citet{sanchez2014}, we have prefered not to correct for the inclination effects in galaxies with inclinations below $35 \degree$, as the uncertainties in the derived correction exceed the very small effect on the spatial distribution of the \hii\,regions.

Finally, the deprojected galactocentric distance for each \hii\,region was then normalised to the disc effective radius ($r_e$), as suggested in \citet{sanchez2012b} and \citet{sanchez2013}. This parameter was determined from the disc scalelength ($r_d$) through the relation \mbox{$r_e = 1.67835 \,r_d$} \citep{sanchez2014}. To obtain the disc scalelength we derived the surface brightness profile of the galaxies by fitting ellipses to the $g$-band light distribution recovered from the data. To do that we used again the {\tt ellipse IRAF} task and fixed the ellipticity and PA values of the successive ellipses matching those of the outer disc (already derived to deproject the galaxies). We then fitted these profiles with the classical exponential decline \citep{freeman1970}, obtaining the value of the corresponding disc scalelength. The values of $r_e$ are given in Table~~\ref{table1} (Appendix~\ref{sec:appendix1}).

\section{Results and discussion}\label{sec:results}

With the methodology explained in the previous section we have obtained the radial distribution of the oxygen abundances for the 102 galaxies in our sample. We describe here the main properties of these abundance profiles and discuss the possible implications of these results on galaxy evolution.

\subsection{The shape of the abundance profiles}\label{sec:shape}

The radial distribution of the gas chemical abundances in spiral galaxies has been extensively studied and, nowadays, it is widely accepted that it presents a negative gradient \citep[e.g.][]{peimbert1979, shaver1983, vilacostas1992, zaritsky1994, kennicutt2003, pilyugin2004, bresolin2009, rosalesortega2011, marino2012, sanchez2012b, pilyugin2014, sanchez2014, sanchezmenguiano2016, zinchenko2016, belfiore2017}. However, recently it has been found that some spiral galaxies show lower (higher) oxygen abundance values in the inner (outer) regions than that expected according to the observed negative gradients. This results in a nearly flat distribution or even a drop towards the centre \citep{belley1992, rosalesortega2011,sanchez2012b,sanchez2014,sanchezmenguiano2016,zinchenko2016}, and an abundance flattening or even an increase in the outermost parts \citep{martin1995, vilchez1996, roy1997, vanzee1998, rosalesortega2011, bresolin2012, marino2012, lopezsanchez2015, sanchezmenguiano2016, belfiore2017}.

In this work we applied a purely automatic procedure (without human supervision) to fit the oxygen abundance radial distribution of the galaxies in the sample, detecting the existence of an inner drop and/or an outer flattening when present. This procedure was designed as follows. The abundance radial distribution of each galaxy is fitted four times, to cover the four possible shapes that have been described in the literature: i) single linear profile, ii) broken linear profile due to the presence of an outer flattening, iii) broken linear profile caused by an inner drop, and iv) doubly-broken linear profile with both an inner drop and an outer flattening. The four mentioned profiles are characterised by these three mathematical functions:

\begin{figure*}
\resizebox{\hsize}{!}{\includegraphics{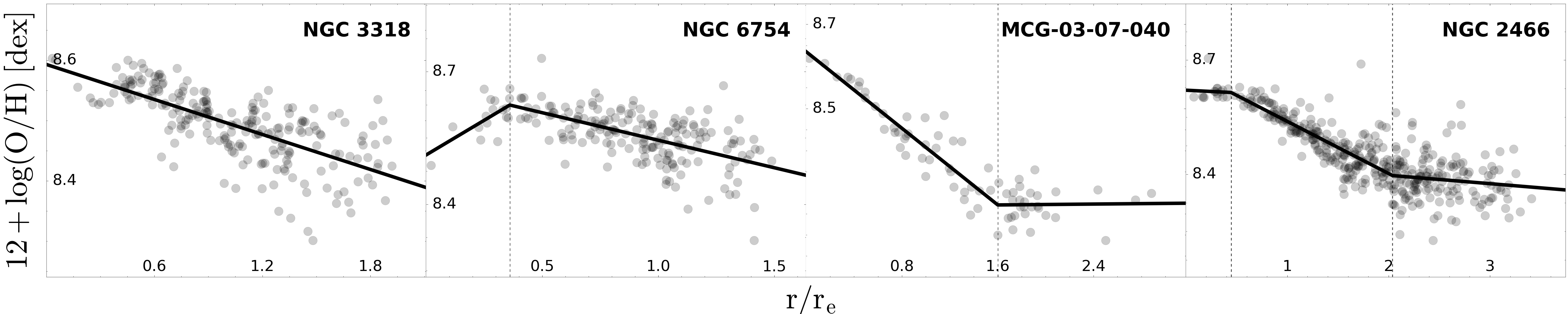}}
\caption{Examples of the different shapes found in the oxygen abundance radial distribution of the \hii\,regions (black markers with transparency). The radial distances are deprojected and normalised to the disc effective radius. The solid black line represents the fit to the distribution. The dashed vertical lines correspond to the radial position of the inner drop and/or outer flattening.}
\label{fig:mosaic}
\end{figure*}

\begin{enumerate}
\item The single profile used to fit the abundance distribution is a linear regression:
\begin{equation}\label{eq1}
O/H \,(r) = a \cdot r + b,
\end{equation}
where $a$ is the slope and $b$ is the zero-point of the linear fit.

\item For the broken profile, the function is constructed by a sum of two linear regressions to be able to fit the principal negative gradient and the possible inner drop or outer flattening, with the radial position of the new feature ($h_i$, with $i=1$ for the inner drop and $i=2$ for the outer flattening) as a free parameter in the fit:
\begin{equation}\label{eq2}
O/H \,(r) = (a_1 \cdot r + b_1) \cdot \left( 1-W_i \right) + (a_2 \cdot r + b_2) \cdot W_i,
\end{equation}
where 
\begin{equation}
W_i = \frac{\pi/2+\arctan\frac{\displaystyle(r-h_i)}{\displaystyle\beta}}{\displaystyle\pi},
\end{equation}
and imposing this condition
\begin{equation}\label{eq3}
a_1 \cdot h_1 + b_1 = a_2 \cdot h_1 + b_2
\end{equation}
to guarantee the continuity of the function, so the oxygen value derived with each linear component is the same at the radial distance at which the consecutive components intersect ($h_i$). For each galaxy this fit is performed twice, to cover the two cases where the broken profile is applicable: negative gradient with an inner drop and gradient with an outer flattening. To distinguish both fits, we imposed lower and upper boundaries on $h_i$ restricting the position of the start of the inner drop to the first half of the covered radial range and the onset of the outer flattening to the second half. In addition, the imposed boundaries avoid that these fits are dominated by just one \hii\,region.
\item For the doubly-broken profile, we introduced another linear component in the sum to fit the cases where both an inner drop and an outer flattening may be present in the galaxy, being the principal negative gradient the component in the middle of both features:
\begin{equation}\label{eq4}
\begin{split}
O/H \,(r) = (a_1 \cdot r + b_1) \cdot \left(1- W_1 \right) + \\ + (a_2 \cdot r + b_2) \cdot W_1 \cdot \left(1-W_2 \right) + (a_3 \cdot r + b_3) \cdot W_2,
\end{split}
\end{equation}
with the following restrictions for continuity:
\begin{equation}\label{eq5}
a_1 \cdot h_1 + b_1 = a_2 \cdot h_1 + b_2
\end{equation}
\begin{equation}\label{eq6}
a_2 \cdot h_2 + b_2 = a_3 \cdot h_2 + b_3.
\end{equation}
\end{enumerate}

The $W_i$ terms are used in the equations to introduce the change of domain among the linear components, acting like a switch: for $r < h_i$, $W_i \simeq 0$ and for $r > h_i$, $W_i \simeq 1$. For instance, in equation \ref{eq2}, for $r < h_i$ we only have contribution of the first term to the fitting function and for $r > h_i$ the whole contribution comes from the second term. For equation \ref{eq4} it is a bit more complicated but the mechanism is the same: the first term of the equation only comes into play for $r < h_1$ and the third one for $r > h_2$. The intermediate term plays a part then between both radial positions ($h_1 < r < h_2$) due to the intersected region where the factors are both contributing (as $W_1=1$ for $r > h_1$ and $1-W_2=1$ for $r < h_2$). This way of constructing the fitting function allows us to find automatically not only the slope and zero-point of all the individual linear components but also the radial position at which the inner drop and outer flattening start using a single and continuous functional form. The $\beta$ parameter was set to a small number to make a very steep transition between consecutive linear components. In this case, we considered $\beta=1\cdot10{^{-8}}''$, although its actual value does not affect the results as long as it is much smaller than $1''$.

Based on the residuals of the four fittings, we selected automatically which was the most suitable one in each case, in the way explained below. We should keep in mind that as soon as we add new components to the fit, the residuals (or scatter) are going to decrease. Therefore, we checked if this reduction in the residuals was statistically significant, making use of a MC simulation. To do that, we generated for each galaxy a `random' radial distribution of the oxygen abundances within the error bars of the measured abundance values, repeating this process for $N = 100$ iterations. We fitted each of these distributions with the four functional forms described before, deriving in all cases the distribution of the residuals of the abundances around the best fitted function. Then, we fitted each of these `residual' distributions with a Gaussian function and we determined if the difference between the mean value derived for the single linear fit (using equation \ref{eq1}) and the mean value for the new fit was higher than the standard deviation value. If this is the case, we can consider that the reduction of the scatter is significant in a statistical way. Finally, to consider this `improved' fitting valid, we imposed another condition: there has to be a significant change in the slope between the consecutive linear components of the corresponding function (equation \ref{eq2} and \ref{eq4}). In order to check this, we fitted again a Gaussian function to the distribution of slopes of the different components, verifying that, similarly to the scatter distribution, the mean values for the inner drop and/or outer flattening were outside the interval ranged between the mean value derived for the principal negative gradient plus or minus the standard deviation.

Figure \ref{fig:mosaic} represents examples of the four different cases that can be found when performing the radial fit (black solid line) of the oxygen abundance distribution of the \hii\,regions (transparent black symbols): single abundance negative gradient (first panel), presence of an inner abundance drop (second panel), presence of an outer abundance flattening (third panel), and presence of both an inner drop and outer flattening (fourth panel). In Appendix~\ref{sec:appendix4} we show the deprojected oxygen abundance radial distribution of the \hii\,regions for the whole sample of galaxies. The slope and zero-point values of the principal negative gradient derived from this fitting, together with the radial distance of the inner drop and/or outer flattening, when present, are shown in Table~\ref{table2} (Appendix~\ref{sec:appendix2}) for all galaxies in the sample.

Applying this procedure with the chosen criteria we can guarantee that the detection of the inner/outer features is reliable. The methodology relies on a sufficient sampling of the \hii\,regions along the covered galactocentric distances for a proper statistical determination of the presence of these features. Otherwise, our procedure seems to stick to the more simple scenario (single slope) and finds better results if those features are neglected. 

In the analysed sample, we find that 55 galaxies exhibit a single abundance gradient. A broken profile due to an inner drop is observed in 21 objects. There are 10 galaxies presenting a broken profile with an outer flattening. Finally, 16 spiral galaxies display a doubly-broken profile in the abundance radial distribution. As we can see, the cases with a clear presence of deviations from the simple negative gradient are as common as the finding of a single behaviour in the radial profiles. This fact suggests that the common view in which the oxygen abundance distribution in spiral galaxies decreases following a single radial gradient might be incomplete and deviations from this single behaviour are needed to be considered for a proper characterisation of the distribution. 

In the case of the abundance profiles derived using the M13-N2 calibrator, we find that: 68 galaxies display a single abundance gradient, 21 of them present a broken profile due to an inner drop, 11 spiral galaxies exhibit a broken profile with an outer flattening, and finally only 2 objects display a doubly-broken profile. For the D16 calibrator, the number of galaxies exhibiting a single abundance gradient increases to 72. The broken profile due the the inner drop is observed in 11, while the broken profile with an outer flattening is observed in 10 objects. Finally, 9 spiral galaxies display a doubly-broken profile. Despite the number of galaxies where an inner drop and/or outer flattening decreases when using the other two calibrators, it can be seen that the presence of deviations from the single behaviour is still remarkable.

Some of these analysed galaxies (7) present peculiarities that can affect the abundance distribution and, although they remain as part of the sample, they were discarded from further analysis: Two galaxies, NGC~232 and NGC~7469, were discarded for having a very powerful AGN in the centre that can contaminate the emission coming from the \hii\,regions (presenting also jets); (ii) another galaxy, NGC~1516A, was removed for being in a very advanced stage of merging; (iii) two more galaxies, UGC~3634 and UGC~6332, were discarded for having a ring-like structure, which affects the accuracy of the derivation of the gradient; (iv) and finally the last two discarded galaxies, NGC~3447 and NGC~7580, are very distorted due to a recent interaction. Therefore, from now on we show results based on a sample of 95 galaxies.

\subsection{Main abundance gradient distribution}\label{sec:slope}

\begin{figure}
\centering
\resizebox{0.95\hsize}{!}{\includegraphics{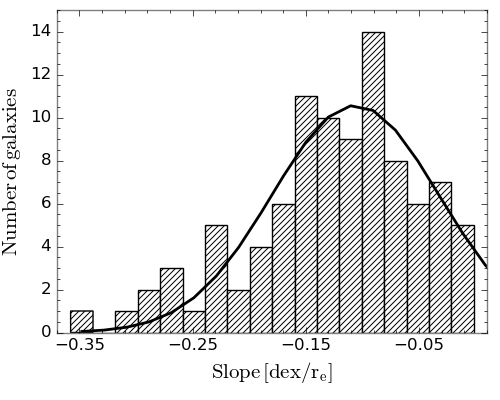}}
\caption{Distribution of slope values of the oxygen main abundance gradients (normalised to the disc effective radius) derived for the galaxy sample. Solid black line represents the Gaussian distribution of the data assuming the mean and standard-deviation of the distribution of slope values and sampled with the same bins.}
\label{fig:slope_dist}
\end{figure}

Figure~\ref{fig:slope_dist} shows the distribution of slopes of the main oxygen abundance gradients (see Sect.~\ref{sec:shape}) derived for the galaxy sample of 95 galaxies. This distribution expands along a wide range of values between approximately $-0.35$ and $0.03$ dex/r$\rm _e$, presenting a clear peak located at $-0.11$ dex/r$\rm _e$. The shape of the distribution suggests the existence of a characteristic gradient more or less similar for all spiral galaxies in the sample. Although using different abundance calibrators and/or different scale lengths to normalise the gradients, several previous works have found this similarity in the slope distribution with a Gaussian function, supporting the existence of the mentioned characteristic gradient \citep{sanchez2012b, sanchez2014, ho2015, sanchezmenguiano2016}. \citet{sanchezmenguiano2016} showed that the slopes of the oxygen abundance gradients in CALIFA spiral galaxies display a distribution peaking at  $-0.07$ dex/r$\rm _e$ ($\sigma=0.05$), in agreement (within the errors) with the value reported in this work. The distribution of slopes for the M13-N2 and D16 calibrators can be found in Fig.~\ref{fig:slopes_others} of Appendix~\ref{sec:appendix3}. In both cases the distributions present a similar peak, but located at $\rm -0.23 \,dex/r_e$ for D16 and at $\rm -0.06 \,dex/r_e$ for M13-N2, reinforcing the idea of a characteristic gradient (although the actual value of the slope changes depending on the used calibrator).

Despite the presence of a clear peak in the distribution of the slopes, the large standard deviation found ($\rm \sigma=0.07 \,dex/r_e$) indicates a possible dependence of such gradients with some particular properties of the galaxies. In this section we will analyse the influence of some characteristics such as the density of the environment, the presence of bars or the type of abundance profile displayed (according to Sect.~\ref{sec:shape}) on the main abundance gradient.

\subsubsection{Influence of the density of the environment}

Numerical simulations have predicted the presence of shallower gradients in interacting galaxies (or galaxies that have undergone an interaction in the past) due to gas flows induced by the interaction \citep[e.g.][]{barnes1996, mihos1996, dalcanton2007, rupke2010a, torrey2012}. This prediction has been confirmed by observational works \citep[e.g.][]{kewley2010, rupke2010b, miralles2014, rosa2014, sanchez2014}.

We classified our galaxies based on the density of their environment to study the possible effect on the abundance gradients. Galaxies were grouped in: isolated (S), part of a group of at least three galaxies (G), in pair (P), and with evidence of real interactions (I). This classification was performed attending to the information found in literature. In Table~\ref{table1} (Appendix~\ref{sec:appendix1}) we show for each galaxy in the sample the corresponding classification and the references used to classify it.

\begin{figure*}
\centering
\resizebox{\hsize}{!}{\includegraphics{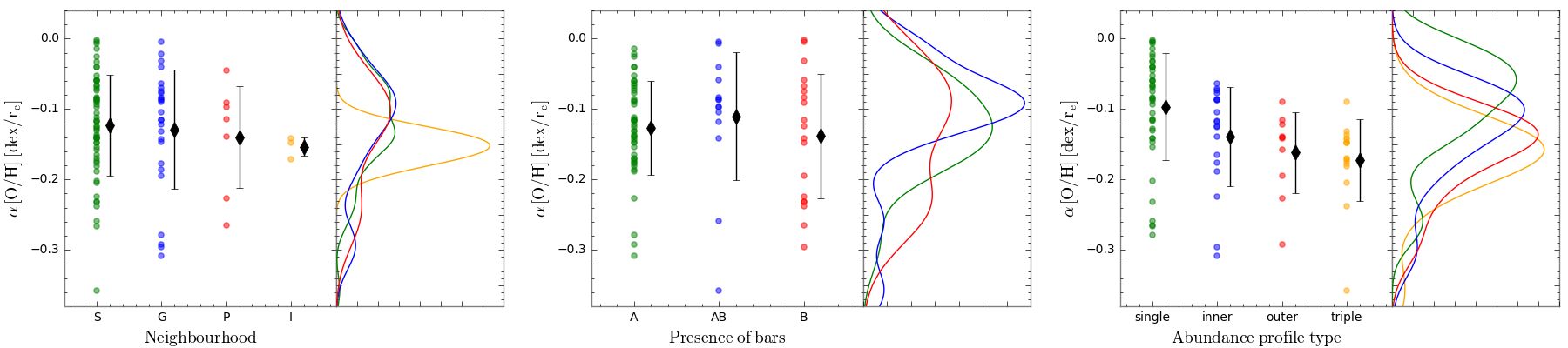}}
\caption{Distribution of slopes depending on: (i) the environment of the galaxies (isolated, green; in groups, blue; paired, red; interacting, yellow) for the {\it left panel}, (ii) the presence or absence of bars (clearly unbarred galaxies, green; clearly barred, red; and an intermediate stage, blue) for the {\it middle panel}, and (iii) the shape of the abundance profile (single profile, green; broken profile with inner drop, blue; broken profile with outer flattening, red; and a doubly-broken profile with presence of both inner drop and outer flattening, yellow) for the {\it right panel}. The black diamonds represent the mean values within each segregation, with the error bars indicating the standard deviation.}
\label{fig:slope_dependences}
\end{figure*}

Figure~\ref{fig:slope_dependences} (left panel) displays the distribution of slope values according to the environment. The black diamonds represent the mean values within each segregation, with the error bars indicating the standard deviation. The mean and standard deviation values of the distribution are the following ones:
\begin{gather*}
{\rm S:}\hspace{8pt} \alpha_{O/H} = -0.12 \,\rm{dex}/r_e \,\,\text{ and }\,\, \sigma = 0.07 \,\rm{dex}/r_e \hspace{12pt}(n_{gal} = 59)\\\relax
{\rm G:}\hspace{6pt} \alpha_{O/H} = -0.13 \,\rm{dex}/r_e \,\,\text{ and }\,\, \sigma = 0.08 \,\rm{dex}/r_e \hspace{12pt}(n_{gal} = 26)\\\relax 
{\rm P:}\hspace{8pt} \alpha_{O/H} = -0.14 \,\rm{dex}/r_e \,\,\text{ and }\,\, \sigma = 0.07 \,\rm{dex}/r_e \hspace{12pt}(n_{gal} = 7)\\\relax 
{\rm I:}\hspace{10pt} \alpha_{O/H} = -0.15 \,\rm{dex}/r_e \,\,\text{ and }\,\, \sigma = 0.01 \,\rm{dex}/r_e \hspace{12pt}(n_{gal} = 3).
\end{gather*}
Although the mean values seem to suggest a trend, the small differences in these values and the high standard deviations found lead us to claim that galaxies present a similar slope independently of the environment where they are. Similar results are found when using the other two calibrators (see Appendix~\ref{sec:appendix3} and Fig.~\ref{fig:slope_dependences_others} therein). However, we have to note that very distorted galaxies due to a recent interaction or to a very advanced stage of merging have been discarded from this analysis since for those objects the definition of the centroid and the disc effective radius are dubious. Therefore, this statement does not necessary contradict previous works claiming the existence of shallower gradients in interacting galaxies. 

\subsubsection{Effects of bars}

Numerical simulations predict that the non-axisymmetrical potential of bars induce radial motions that can homogenise the gas producing shallower gradients \citep{friedli1994, friedli1995, cavichia2014}. On the other hand, observations diverge in their conclusions. Some works have found evidence of flatter gradients in barred galaxies with respect to unbarred systems \citep{vilacostas1992,martin1994, zaritsky1994, dutil1999}. However, recent studies based on larger statistical samples have not found evidence of such correlation between the slope of the abundance gradient and the presence of a bar \citep{sanchez2014, sanchezmenguiano2016}.

In order to test these results, we define three different groups according to the presence or absence of bars: galaxies with no bar (A), galaxies that may have a bar but it is not clearly visible (AB), and clearly barred galaxies (B). This information was collected from the extragalactic database {\it HyperLeda} \citep{makarov2014}. The galaxies for which such information is not published (16) were not considered in this comparison, but were not excluded from the rest of the analysis.

Figure~\ref{fig:slope_dependences} (middle panel) shows the distribution of slopes according to the presence of bars. The black diamonds represent the mean values within each group, with the errorbars indicating the standard deviation. These values are:
\begin{gather*}
{\rm A:}\hspace{13pt} \alpha_{O/H} = -0.13 \,\rm{dex}/r_e \,\,\text{ and }\,\, \sigma = 0.07 \,\rm{dex}/r_e \hspace{10pt}(n_{gal} = 46)\\\relax
{\rm AB:}\hspace{6pt} \alpha_{O/H} = -0.11 \,\rm{dex}/r_e \,\,\text{ and }\,\, \sigma = 0.09 \,\rm{dex}/r_e \hspace{10pt}(n_{gal} = 14)\\\relax
{\rm B:}\hspace{13pt} \alpha_{O/H} = -0.14 \,\rm{dex}/r_e \,\,\text{ and }\,\, \sigma = 0.09 \,\rm{dex}/r_e \hspace{10pt}(n_{gal} = 19). 
\end{gather*}
We can see that unbarred (green) and barred galaxies (blue and red) present a similar distribution of slopes with very similar mean values. This suggests a lack of correlation between the abundance slope and the barred/unbarred nature of galaxies, in agreement with \citet{sanchez2014} and \citet{sanchezmenguiano2016}. This lack of correlation also appears when using the other two calibrators (see Appendix~\ref{sec:appendix3} and Fig.~\ref{fig:slope_dependences_others} therein).

\subsubsection{Dependence on the shape of the abundance profile}

The existence of different types of abundance profiles in the galaxy sample is associated with the presence of deviations from the most commonly reported behaviour in the abundance radial distribution. An interesting point to investigate would be the effect of the presence of these deviations in the main abundance gradient. To this end, we show in the right panel of Fig.~\ref{fig:slope_dependences} the dependence of the slope in the negative gradient with the shape of the abundance profile: a single gradient (green), a broken linear profile due to the inner drop (blue) and to the outer flattening (red), and a doubly-broken linear profile due to the presence of both features (yellow). For this analysis we have discarded 10 galaxies not presenting signs of the presence of the outer flattening but for which the MUSE FoV does not cover the full disc extent, and therefore may display this feature and we are not able to detect it.

The slope distributions represented in the right panel of Fig.~\ref{fig:slope_dependences} show a trend in the slope with the shape of the abundance profile, with the single profiles presenting the shallowest gradients and the doubly-broken profiles the steepest ones. The mean and standard deviation values (in units of dex/r$_{\rm e}$) of the distributions are:
\begin{gather*}
{\rm Single:}\hspace{48.5pt} \alpha_{O/H} = -0.10 \,\text{ and }\, \sigma = 0.07 \hspace{5pt}(n_{gal} = 42)\\\relax
{\rm Broken \,(inner):}\hspace{15.5pt} \alpha_{O/H} = -0.14 \,\text{ and }\, \sigma = 0.07 \hspace{5pt}(n_{gal} = 19)\\\relax 
{\rm Broken \,(outer):}\hspace{15.5pt} \alpha_{O/H} = -0.16 \,\text{ and }\, \sigma = 0.06 \hspace{5pt}(n_{gal} = 10)\\\relax 
{\rm Doubly-broken:}\;\; \alpha_{O/H} = -0.17 \,\text{ and }\, \sigma = 0.06 \hspace{5pt}(n_{gal} = 16).
\end{gather*}
These results indicate that the presence of the inner drop and the outer flattening does have an effect on the negative abundance gradient displayed by spiral galaxies. This general trend is also observed in the cases where the M13-N2 and D16 calibrators are used (see Fig.~\ref{fig:slope_dependences_others}). For the M13-N2 (D16) indicator, we find a total difference of $-0.04 \,(-0.09)$ dex/r$_{\rm e}$ between the galaxies presenting a single abundance gradient and the ones displaying a doubly-broken profile (see Appendix~\ref{sec:appendix3}).

In previous works analysing galaxies from the CALIFA survey \citep{sanchez2014,sanchezmenguiano2016} we already found slightly steeper gradients in galaxies with evidence of an inner drop. The fact that the presence of this feature had an impact on the overall distribution of abundances at larger radii led us to propose the existence of gas radial motions \citep{lacey1985, portinari2000, schonrich2009, spitoni2013} as the cause of this feature. These radial movements would be towards the knee point in the abundance distribution (drop), outwards in the inner regions and inwards in the outer parts. In this work, using a different sample of galaxies and a different dataset, we confirm these results and find a similar situation regarding the presence of the outer flattening. Different mechanisms have been suggested to explain the origin of the outer flattening in the gas abundances, being radial motions of both gas and stars one of them \citep[e.g.][]{goetz1992, ferguson2001, sellwood2002, minchev2010, bilitewski2012, rovskar2012, daniel2015}. On the other hand, alternative mechanisms such as minor mergers and satellite accretion \citep{quillen2009,qu2011,bird2012}, a radial dependence of the star formation efficiency at large galactocentric distances \citep{bresolin2012,esteban2013} or a balance between outflows and inflows with the intergalactic medium \citep{oppenheimer2008, oppenheimer2010, dave2011, dave2012} have also been proposed. All these mechanisms are not mutually exclusive and to date we have not been able to distinguish between them. However, although we do not discard the other mechanisms to also be involved, the results found in this work suggest that radial motions play a fundamental role shaping the abundance profiles of spiral galaxies by producing an abundance drop in the inner regions and/or an abundance flattening in the outermost parts. A deeper study on the connection of these features with radial motions, also from the dynamical and theoretical point of view, might help to further constrain this scenario.

\subsection{Location of the inner drop and the outer flattening}\label{sec:dropflattening}

As stated previously, the presence on an inner drop and an outer flattening in the radial abundance profiles of some spiral galaxies was already found in several works. However, the few of them that analysed these features in a statistical way using a large sample of galaxies \citep{sanchez2014, sanchezmenguiano2016} always assumed that their location was the same for all the galaxies where they were observed. This assumption was based on a visual inspection of the sample, and afterwards confirmed by the stacking of the abundance profiles of the individual galaxies. 

To date, no previous studies have carried out a detailed analysis of these features, determining the actual location of the inner drop and the outer flattening for a large sample of galaxies. Therefore, here we have developed a methodology to fit the radial abundance profiles in an automatic way that allows us to derive the location of these features. Hereafter, we will define $h_D$ as the galactocentric distance at which the inner drop appears ($h_1$ in equations from \ref{eq2} to \ref{eq6}) and $h_F$ as the distance at which the radial gradients flatten ($h_2$ in equations from \ref{eq2} to \ref{eq6}).

\begin{figure}
\centering
\resizebox{0.95\hsize}{!}{\includegraphics{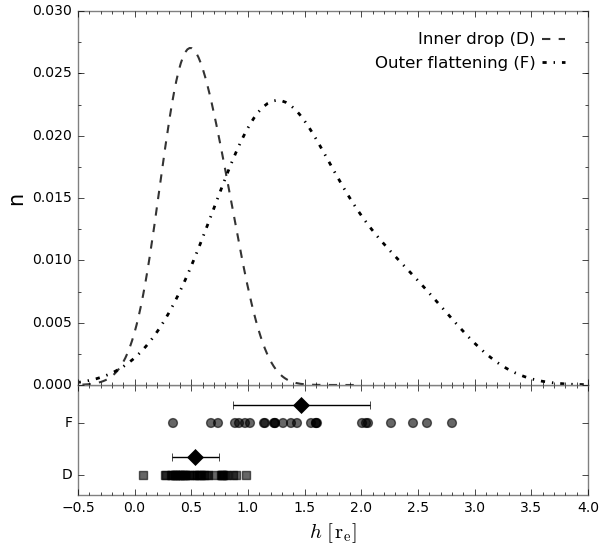}}
\caption{Distribution of radial positions (in units of $\rm r_e$) of the inner drop (D, squares) and outer flattening (F, dots) found in some galaxies of the sample. The black diamonds represent the mean values of the position at which these features appear, with the error bars indicating the standard deviations. Dashed and dashed-dotted lines represent the normalised density distributions of the location of the inner drop and outer flattening, respectively.}
\label{fig:dropflat}
\end{figure}

In Fig.~\ref{fig:dropflat} we represent the distribution of values derived for $h_D$ (dashed line) for the 37 galaxies where the inner abundance drop is detected and the distribution of values of $h_F$ for the 26 galaxies displaying the outer abundance flattening  (dashed-dotted line). These density (normalised to one) distributions were derived considering each radial position as a Gaussian distribution centred at such point with a sigma given by the maximum distance between nearest-neighbours, and summing all these individual Gaussians. In this way we reproduce a smooth distribution that is more peaky in the clustered points. As it can be seen in the figure, the distribution of the radial positions of the inner drop is quite narrow and has a pronounced peak centred at $\rm \sim 0.5 \,r_e$ (the mean value of the radial positions is 0.54 r$\rm _e$ and the standard deviation 0.20), suggesting that the position of the inner drop is very similar in all galaxies where this feature is present. However, the distribution for the outer flattening is much wider, with the flattening occurring between 0.3 and 2.8 r$\rm _e$, and overlapping the distribution of the inner drop. Although the distribution shows a peak at $\rm \sim 1.5 \,r_e$ (the mean value of the radial positions is 1.47 r$\rm _e$ and the standard deviation 0.60), the large range of radii covered by it prevents us to confirm a characteristic location of the outer flattening in the abundances.

The use of the other two calibrators reduces the number of galaxies where the inner abundance drop is detected to 23 for M13-N2 and to 20 for D16. The number of galaxies displaying the outer abundance flattening is also reduced to 13 galaxies in the case of the M13-N2 calibrator and to 19 in the case of D16. However, the distribution of the radial positions of both features is very similar to the one obtained with O3N2 (see Fig.~\ref{fig:dropflat_others}), with very slight differences in the average positions (inner drop at $\rm \sim 0.8 \,r_e$ for M13-N2 and at $\rm \sim 0.6 \,r_e$ for D16; outer flattening at $\rm \sim 1.5 \,r_e$ and at $\rm \sim 1.6 \,r_e$, respectively).

These results are in agreement with previous works assuming a similar location of the inner drop (0.5 r$\rm _e$) for all galaxies displaying this feature in the radial abundance distribution \citep{sanchez2014, sanchezmenguiano2016b}. However, our results seem to contradict the claim that this was also the case for the outer flattening (located approximately at $\rm \sim 2.0 \,r_e$). These works were all based on visual inspections of the general shape of the gradient (see Sect.~\ref{common} below), placing the location of the outer flattening at the position in which the gradient clearly deviates from its linear behaviour (that would happen approximately between $2-2.25\,r_e$ according to Fig.~\ref{fig:commongrad}). Indeed, this location has fluctuated between different estimations, being placed at $2.2\,r_e$ by \citet{sanchez2012b} and around 2 by \citet{sanchez2014}. However, the detailed analysis carried out in this work indicates that the knee point is nearer to $\rm \sim 1.5 \,r_e$, with a large dispersion of values among galaxies. Thus, our results do not imply a contradiction, but reflects the need of a more detailed analysis like the one presented here.

\subsection{A common oxygen abundance gradient}\label{common}

\begin{figure}
\centering
\resizebox{\hsize}{!}{\includegraphics{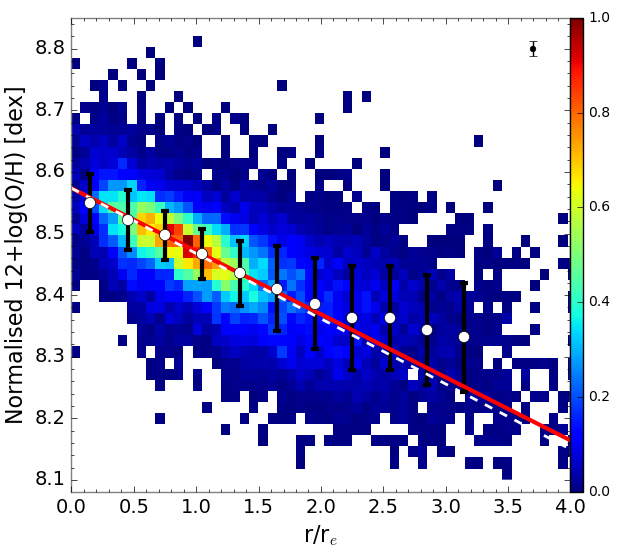}}
\caption{Radial density distribution of the oxygen abundance after normalising the abundances of each galaxy so that the abundance at $\rm 1 r_e$ is equal to the average value for the whole sample at such radius. The white dots represent the mean oxygen abundance values, with the error bars indicating the corresponding standard deviations, for bins of 0.3 $\rm r_e$. The solid red line represents the error-weighted linear fit derived for those mean values within the range between 0.5 and 1.5 $\rm r_e$, and the dashed white line represents the linear relation corresponding to the characteristic value of the slope derived in Sect.~\ref{sec:slope} for the individual galaxies assuming a Gaussian distribution.}
\label{fig:commongrad}
\end{figure}

The existence of a characteristic slope in the oxygen abundance distribution of spiral galaxies was stated in \citet{sanchez2012b,sanchez2014} and later confirmed in \citet{sanchezmenguiano2016}. In this work, following a similar recipe, we derived the radial distribution of the oxygen abundances for the 95 galaxies in the sample and stacked all of them as a single distribution. To do that we rescaled the distributions, as the $\mathcal{M}-\mathcal{Z}$ relation \citep{tremonti2004} produces an offset of the abundance profile according to the integrated stellar mass of the galaxy. To this end, we applied an offset to the distribution of each galaxy by normalising the abundance at the disc effective radius to the average value for the whole sample ($\sim 8.5$ dex). The final outcome of this analysis is shown in Fig.~\ref{fig:commongrad}. The abundance distribution is represented as a colour-coded density map (normalised to one) and shows clearly a common abundance gradient presented by all galaxies in the sample. The white dots represent the mean oxygen abundance values, with the error bars indicating the corresponding standard deviations, for bins of 0.3 $\rm r_e$. An error-weighted linear regression (solid red line) to the mean values within the range between 0.5 and 1.5 $\rm r_e$ (trying to avoid the regions dominated by the presence of the inner abundance drop and outer flattening) derives a slope of $\alpha_{O/H} = -\,0.10\pm0.03\,\rm{dex}/r_e$. This is compatible with the characteristic slope of $\alpha_{O/H} = -\,0.11 \,(\sigma = 0.07) \,\rm{dex}/r_e$ derived in Sect.~\ref{sec:slope} for the individual galaxies assuming a Gaussian distribution (dashed-white line).

As with M13-N2 and D16, the common abundance gradient is clearly visible when representing the radial abundance distribution of the whole sample (with a slope of \mbox{$\alpha_{O/H} = -0.04$} and \mbox{$-0.20 \,\rm{dex}/\rm{r_e}$} for M13-N2 and D16, respectively). However, the presence of the inner drop and the outer flattening is not so evident. This is somehow expected due to the wide range of values found for $h_D$ and $h_F$ that may blur their signatures when all gradients are stacked together. 

\subsection{The characteristic slope as a normalisation scale}\label{masses}

Recently, \citet{zinchenko2016} showed the validity of the use of gas abundance maps to determine the geometrical parameters of galaxies (coordinates of the centre, inclination and position angle of the major axis). Generally, these parameters are derived from the photometric analysis of a galaxy, assuming that the surface brightness of the galaxy is axisymmetric. However, if we consider that the metallicity of the disc is also a function of the galactocentric distances \citep[i.e. the possible azimuthal asymmetries, in case of existing, are small, e.g.][]{sanchezmenguiano2017}, then we can expect that the abundance map can also be used for the determination of geometrical parameters. 

\begin{figure}
\centering
\resizebox{\hsize}{!}{\includegraphics{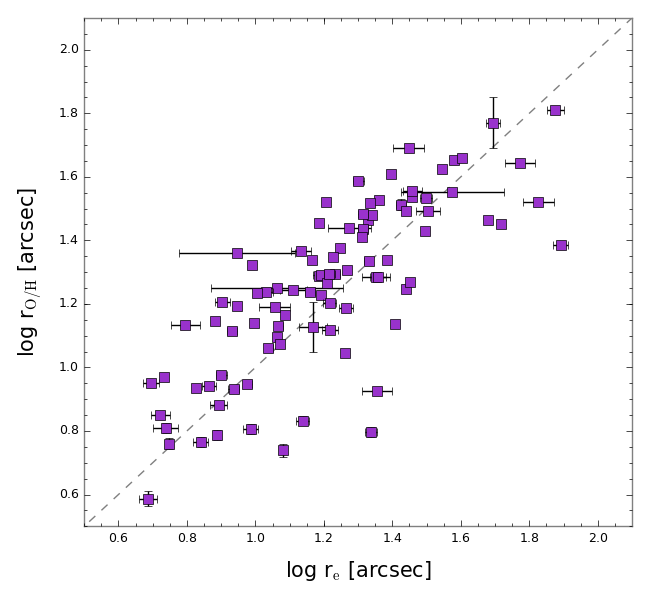}}
\caption{Radius at which the abundance distribution decays 0.10 dex (given by the decrease in the abundance in one disc effective radius attending to the common abundance gradient, see Fig.~\ref{fig:commongrad}) versus the disc effective radius (in logarithmic scale).}
\label{fig:radrelation}
\end{figure}

The results presented in \citet{zinchenko2016} could be a consequence of the oxygen abundance distribution following the surface mass density and therefore the surface brightness of a galaxy, as a result of the local $\Sigma-\mathcal{Z}$ relation \citep{rosalesortega2012,sanchez2013, barreraballesteros2016}. Due to this connection between both properties, we could define a new scale length based on the abundance maps to normalise the abundance profiles instead of using the disc effective radius. In the previous section we confirmed the presence of a characteristic slope of $-0.10\,\rm{dex}/r_e$ in the abundance distribution. Therefore, we define here a parameter named `the abundance scale length' ($r_{O/H}$) as the radial position at which the abundance distribution of a galaxy decays $0.10\,\rm{dex}$. If we represent $r_{O/H}$ as a function of $r_e$, we would expect the distribution of values falling near the 1:1 relation, indicating that the surface brightness and the gas metallicity follow the same distribution. The outcome of this statement is shown in Fig.~\ref{fig:radrelation}, obtaining a tight correlation between both parameters. The values of $r_{O/H}$ for all galaxies in the sample are shown in Table~\ref{table2} (Appendix~\ref{sec:appendix2}).

\begin{figure}
\centering
\resizebox{0.95\hsize}{!}{\includegraphics{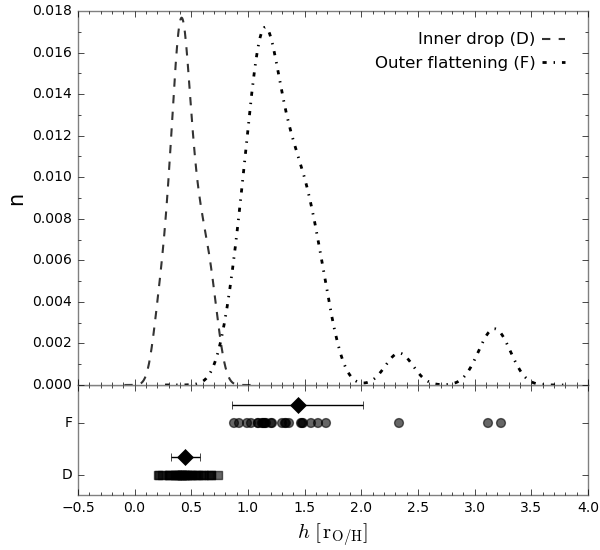}}
\caption{Same as Fig.~\ref{fig:dropflat} but normalising the abundance distribution to r$_{O/H}$. See caption above for more details.}
\label{fig:dropflatralpha}
\end{figure}

We normalise now the radial abundance distributions of the galaxy sample using $r_{O/H}$ instead of $r_e$ as previously done, and repeat part of the analysis. The new distributions of radial positions of the inner drop and the outer flattening is shown in Fig.~\ref{fig:dropflatralpha} (specific values are listed in Table~\ref{table2}, Appendix~\ref{sec:appendix2}). It can be seen that now the distributions have got significantly narrower, especially in the case of the outer flattening (with the exception of three outliers). The mean values of the radial position of both features are very similar to the derived ones normalising to $r_e$ ($h_D = 0.45 \:{\rm r_{O/H}}$ and $h_F = 1.44 \:{\rm r_{O/H}}$). However, the standard deviations around the mean values have been reduced more than a half of the previous value (from 0.20 to 0.13 for the inner drop and from 0.60 to 0.20 for the outer flattening, not considering the three outliers). This is also the case when using the M13-N2 and the D16 calibrators (see Fig.~\ref{fig:dropflatralpha_others}), although the reduction of the standard deviation values is smaller than for the M13-O3N2 calibrator (from 0.22 to 0.17 and from 0.22 to 0.16 for the inner drop, from 0.53 to 0.22 and from 0.55 to 0.35 for the outer flattening, using the M13-N2 and the D16 calibrators, respectively).

\begin{figure*}
\resizebox{\hsize}{!}{\includegraphics{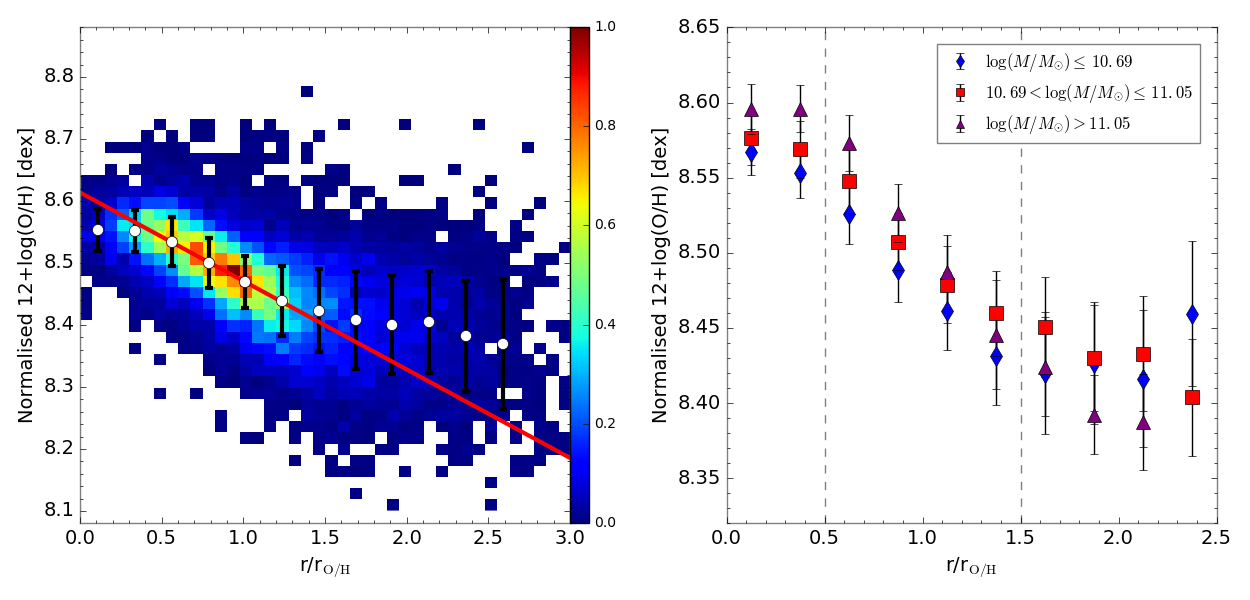}}
\caption{{\it Left}: Same as Fig.~\ref{fig:commongrad} but normalising the abundance distribution to r$_{O/H}$. See caption above for more details. {\it Right}: Mean oxygen abundance radial profiles derived for galaxies belonging to three different stellar mass bins: $\log (M/M_\odot) \leq 10.69$, blue diamonds; $10.69 < \log (M/M_\odot) \leq 11.05$, red squares; $\log (M/M_\odot) > 11.05$, purple triangles. The limits of the bins were chosen to ensure a similar number of elements in each bin.  The symbols represent the mean oxygen abundance values, with the error bars indicating
the corresponding standard deviations, for bins of 0.25 r$_{O/H}$. Dashed vertical lines indicate the average position of the inner drop and the flattening in the outer parts.}
\label{fig:commongrad_ralpha}
\end{figure*}

Finally, we reproduce Fig.~\ref{fig:commongrad} stacking all the individual abundance distributions of the galaxy sample but using now the new normalisation scale, $r_{O/H}$. The result is shown in the left panel of Fig.~\ref{fig:commongrad_ralpha} (similar figures for the M13-N2 and the D16 calibrators are shown in Appendix~\ref{sec:appendix3}, Fig.~\ref{fig:commongrad_ralpha_others}). By construction, the global abundance distribution presents a characteristic slope of $-0.10\,\rm{dex}/r_{O/H}$ ($-0.20\,\rm{dex}/r_{O/H}$ in the case of the M13-N2 calibrator and $-0.04\,\rm{dex}/r_{O/H}$ for D16). In addition, it is evident now the presence of the inner drop and the outer flattening in the abundances, features that were blurred when normalising the abundance distribution to $r_e$. This is because now these features happen at a similar radial position for the whole sample, so regions presenting different behaviours have not been mixed when stacking the distribution of all the galaxies.

The right panel of Fig.~\ref{fig:commongrad_ralpha} shows the average oxygen abundance radial profiles when separating the sample into different bins according to the integrated stellar mass of the galaxies (similar figures for the M13-N2 and D16 calibrators are shown in Appendix~\ref{sec:appendix3}). It can be seen that the presence of the abundance drop in the inner regions of the galaxy discs and the flattening in the outer regions is also easily observable (although for D16 the presence of the inner drop is less evident). We calculated the stellar masses using the mass-luminosity ratio presented in \citet{bell2001}. We derived the $g$ and $r$ apparent magnitudes from the flux density in the $g$- and $r$-band images recovered from the data and we used the equations presented in \citet{jester2005} to transform SDSS magnitudes to the Johnson system. Then we transformed these quantities to luminosities knowing the galaxy distance. Finally, from the $V$-band luminosity and the $\rm B-V$ colour we obtained the stellar masses using the mentioned mass-luminosity ratio. The luminosity and colour values were corrected by dust attenuation. The integrated values were determined from the values derived at one disc effective radius since some galaxies are not fully covered by the FoV of MUSE. We performed a sanity check to test the robustness of the derived stellar masses (with integrated values determined from the values at one disc effective radius) by comparing them with the values derived using directly the integrated values in the cases where the extension of the galaxies was completely covered by the FoV of MUSE. The correlation was very good, with a standard deviation of just 0.12 dex, similar to the typical systematic errors on the mass derivation.

\begin{figure}
\centering
\resizebox{0.95\hsize}{!}{\includegraphics{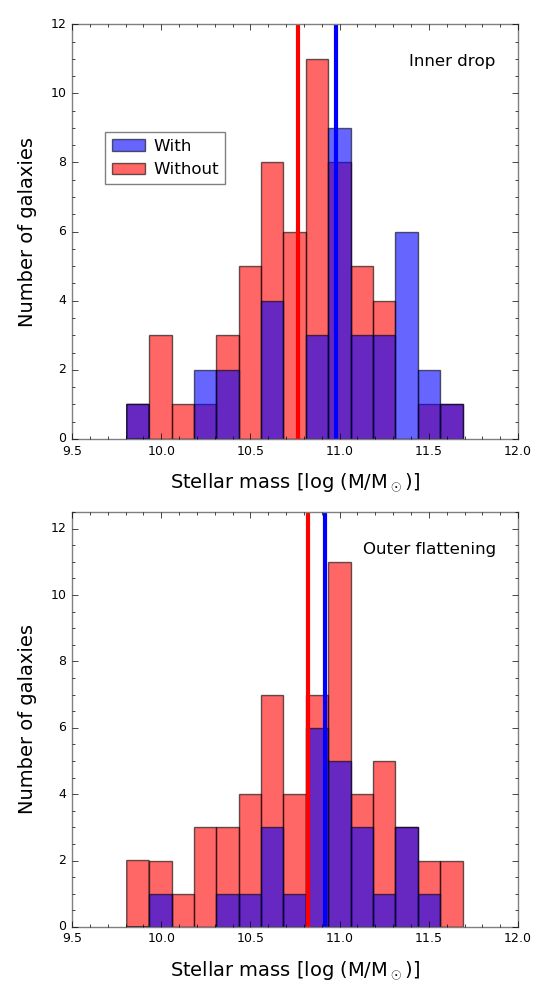}}
\caption{Distribution of the integrated stellar mass of the galaxy sample for galaxies with (blue) and without (red) presence of an abundance drop in the inner regions of the galaxy discs ({\it top panel}) and an abundance flattening in the outer regions ({\it bottom panel}). Vertical lines represent the mean values for each distribution.}
\label{fig:masstypes}
\end{figure}

The limits of the mass bins were chosen to ensure a similar number of elements in each bin: $\log (M/M_\odot) \leq 10.69$ (blue diamonds), $10.69 < \log (M/M_\odot) \leq 11.05$ (red squares), $\log (M/M_\odot) > 11.05$ (purple triangles). In \citet{sanchezmenguiano2016} we found a dependence of the presence of the inner drop with the integrated stellar mass of the galaxies, with the most massive galaxies presenting the deepest abundance drops. The flattening on the opposite appeared independently of the galaxy mass. Similar trends are obtained in this work, confirming the result reported in \citet{sanchezmenguiano2016}. To explore this result, we represent in Fig.~\ref{fig:masstypes} the distribution of the integrated stellar mass for the galaxies with (blue histograms) and without (red histograms) presence of the abundance drop in the inner regions (top panel) and the outer flattening (bottom panel). Again, in the analysis of the outer flattening we have discarded the 10 galaxies for which we do not detect an outer flattening maybe due to an incomplete coverage of the disc extent by the MUSE FoV. It can be seen that galaxies presenting an inner drop are more massive than the galaxies without evidence of this feature. The mean stellar mass for the former ones (blue vertical line) is $10.98 \pm 0.07$ ($\log (M/M_\odot)$), whereas in the latter the mean value (red vertical line) is $10.77 \pm 0.05$ ($\log (M/M_\odot)$). Regarding the outer flattening, the distributions for the galaxies with and without presence of this feature cover the same range of masses, with similar mean values ($10.92 \pm 0.06$ for galaxies with outer flattening and $10.82 \pm 0.06$ for galaxies with no presence of this flattening). A KS-test yields a p-value value of $1\%$ when comparing the distributions with and without an inner abundance drop, while we obtain a p-value of $40\%$ when analysing the distributions with respect to the presence of the outer flattening. Therefore, this test also confirms the results reported in \citet{sanchezmenguiano2016}, that is, there is a dependence of the presence of the inner drop with the galaxy mass, and an absence of such relation in the case of the outer flattening.

\subsection{Dispersion of the gradient and mixing scalelength}\label{sec:scatter}

Only a small number of studies focussed on individual galaxies have analysed the scatter relative to the radial abundance gradients, searching for evidences of azimuthal variations in the metallicity. Analysis on NGC~300 \citep{bresolin2009b} or M~33 \citep{bresolin2011} reveal small scatter values consistent with the uncertainties in the abundance measurements. On the other hand, recent observational studies on NGC~6754 \citep{sanchez2015a} and M~101 \citep{croxall2016}, although based on very different approaches to derive the oxygen abundance of the gas component, have both found a substantial dispersion that can be taken as an evidence of the presence of azimuthal variations. The results on NGC~6754 were later confirmed by \citet{sanchezmenguiano2016b} by analysing the azimuthal abundance distribution of the galaxy. However, regarding M~101, a previous study performed by \citet{li2013} only observed marginal local metallicity inhomogeneities that the authors suggested may be linked to tidal features reported at that time \citep{mihos2012}. 

Due to the large number of \hii\,regions needed to perform a proper analysis of the dispersion of the abundance gradients, an extended study on this topic has not been feasible to date, being limited only to individual galaxies. The number of galaxies in our sample in which a large number of \hii\,regions is detected due to the high resolution of MUSE data makes this analysis perfectly achievable in this work. Studying the possible relation between the dispersion and other parameters of the galaxies such as the presence of a bar, the morphological type or the stellar mass can help us to constrain the possible factors that can produce a significant scatter in the abundance gradients.

\begin{figure*}
\begin{center}
\resizebox{\hsize}{!}{\includegraphics{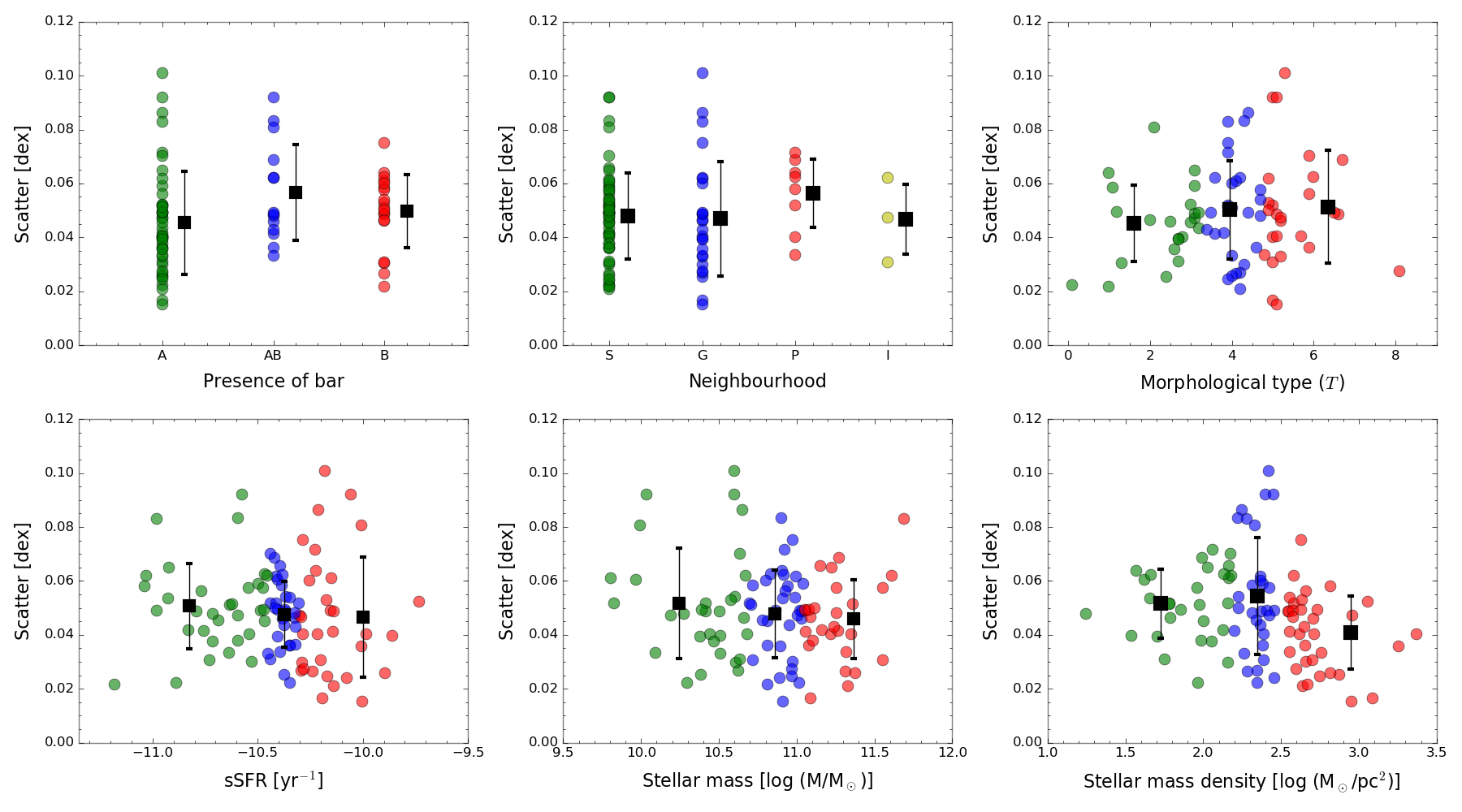}}
\caption{Distribution of the scatter relative to the radial abundance gradients as a function of: presence of bars (A, clearly unbarred galaxies; B, clearly barred; AB, intermediate stage; {\it top left}), environment of the galaxy (S, isolated; G, in group; P, paired, I, in interaction; {\it top middle}), numerical morphological type according to the de Vaucouleurs system (T, {\it top right}), specific star formation rate ({\it bottom left}), integrated stellar mass ({\it bottom middle}), and stellar mass density at one disc effective radius ({\it bottom right}). The black squares represent the mean values within each segregation, with the errorbars indicating the standard deviation. See text for details of the derivation of these parameters.}
\label{fig:scatter}
\end{center}
\end{figure*}

With this purpose we measured the scatter relative to the radial abundance gradient of each galaxy of the sample as the standard deviation of the differences between the observed abundances and the corresponding values from the linear regression according to their deprojected galactocentric distances. We only derived the scatter in the radial range covered by the linear behaviour (see Sect.~\ref{sec:shape} for more details about this linear regime). Figure~\ref{fig:scatter} represents the distribution of scatter values according to different characteristics of the galaxies: presence of bars (A, clearly unbarred galaxies; B, clearly barred; AB, intermediate stage; top left), environment of the galaxy (S, isolated; G, in group; P, paired, I, in interaction; top middle), numerical morphological type according to the de Vaucouleurs system from {\it HyperLeda} (T, top right), specific star formation rate (bottom left), integrated stellar mass (bottom middle), and stellar mass density at one disc effective radius (bottom right). The values of these parameters for all galaxies in the sample are given in Table~\ref{table1} (Appendix~\ref{sec:appendix1}), except the scatter values that are provided in Table~\ref{table2} (Appendix~\ref{sec:appendix2}). The star formation rate (SFR) was derived for each galaxy adopting the classical relation presented in \citet{kennicutt1998} based on the dust-corrected H$\alpha$ luminosities. These luminosities were obtained deriving the apparent magnitudes from the flux density in an H$\alpha$ image recovered from the data and then transforming these quantities to luminosities knowing the galaxy distance. The specific star formation rate (sSFR) for each galaxy was then obtained dividing the SFR by the mass (see Sect.~\ref{masses} for details on the derivation of the galaxy mass). The relations between the scatter and the different properties represented in Figure~\ref{fig:scatter} show no significant dependence of the scatter with any of the analysed parameters. Indeed, the measured values are in general lower than the error of the calibrator used to derive the oxygen abundances ($\sim0.08$ dex), and therefore, we cannot conclude that the detected scatter has a physical origin. 

Due to the reduction of the physical spatial resolution when we move to high redshift, we observe that the scatter increases with the number of the detected \hii\,regions in a galaxy (as the number of these regions decreases with the redshift). In order to avoid this non-physical bias we have discarded those galaxies with $z > 0.02$ and less than 30 \hii\,regions detected. Reproducing the analysis with the remaining 52 galaxies, we obtain similar results, that is, there is no dependence of the scatter relative to the radial abundance gradient with any of the analysed properties of the galaxies.

The dispersion around the radial abundance gradient of a galaxy can be considered an upper limit to the maximum radial mixing scale, as any radial mixing increases the scatter by moving regions of a certain abundance from one galactocentric distance to another \citep[e.g.][]{scalo2004}. If we divide the scatter by the slope of the gradient we obtain a measurement of how far a certain \hii\,region has been able to move from its expected location based on the displayed abundance gradient, that is, the mixing scalelength \citep[$r_{mix},$][]{sanchez2015a}. The derived values for our sample typically range between 0.1 and 0.8 r$\rm _e$, with a median value of 0.4 r$\rm _e$, very similar to the mixing scalelength of 0.43 r$\rm _e$ derived for NGC~6754 by \citet{sanchez2015a}. As in the case of the scatter, we find no significance dependence of the mixing scalelength with any of the analysed properties of the galaxies.


\section{Conclusions}\label{sec:conclusions}

The existence of a radial decrease in the gas chemical abundances of spiral galaxies was well established decades ago \citep{searle1971, comte1975, smith1975}, supporting the inside-out scenario for disc evolution. However, in recent years there have been a large number of works showing observational evidence of the existence of some deviations from this pure negative gradient. This way, the presence of a decrease of the abundance in the inner regions and a flattening in the outer parts of some spiral galaxies is nowadays widely known \citep[e.g.][]{belley1992, vilchez1996, bresolin2009, rosalesortega2011, sanchez2014, sanchezmenguiano2016}. Despite this, we still lack of a good characterisation of these features and, therefore, of the radial abundance distribution itself. The advent of MUSE, an instrument combining a large FoV with high spatial resolution, can help us to better characterise the radial abundance distribution of spiral galaxies, allowing us to increase the number of \hii\,regions detected in individual galaxies with respect to previous studies. The high spatial resolution also helps to reduce the contamination of the diffuse emission in the detected regions and the dilution effects. All these advantages result in a better mapping of the radial distribution, including the inner and outermost parts where deviations from the linear behaviour appear.

In this work we characterised the oxygen abundance radial profile (normalising it to the disc effective radius) of a large sample of 102 spiral galaxies using MUSE data, allowing us to obtain statistically significant results. We developed a new methodology to automatically fit the radial profiles, detecting the presence of possible deviations in the radial distribution with respect to the linear behaviour in the inner and outer parts of the discs (i.e. the inner drop and the outer flattening). We found 55 galaxies presenting a single abundance gradient, 37 galaxies exhibiting signs of an inner drop in the abundances, and 26 galaxies showing a flattening in the outermost part of the discs (a galaxy can present both features simultaneously). These numbers reflect that the existence of these features in the radial abundance distribution is very common, leaving behind the scenario where the oxygen abundance distribution of spiral galaxies was well described only by a radial negative gradient. 

We derived the distribution of slopes for the abundance gradient in a sub-sample of 95 galaxies (after discarding 7 galaxies that present some peculiarities that could affect the shape of the radial abundance distribution). We find a clear peak in the distribution suggesting the existence of a characteristic gradient in spiral galaxies \citep[first revealed by][]{sanchez2014}. If we represent the radial abundance distribution of the whole sample together (applying a different offset to each of the individual gradients to take into account the effects of the $\mathcal{M}-\mathcal{Z}$ relation), this characteristic gradient is clearly evident, with a slope of $\alpha_{O/H} = -\,0.10\pm0.03\,\rm{dex}/r_e$. 

The large dispersion found in the distribution of slopes could indicate a possible dependence of the gradient with some particular property of the galaxies. Analysing the effects of the density of the environment and the presence of bars, no significant differences have been found in the slope. However, we find steepest gradients when the presence of an inner drop or an outer flattening is detected, suggesting that radial motions might be playing an important role shaping the abundance profiles and causing these features. Although future studies are needed to understand these motions and their origin, we are setting important observational constraints for theoretical works on radial migration.

The methodology developed in this work allowed us to study the radial position of the inner drop and the outer flattening. To our knowledge, no previous works have carried out this type of analysis. We find that the inner drop appears at a very similar galactocentric distance for all the galaxies (approximately $\rm 0.5 \,r_e$), while the distribution of the radial positions of the outer flattening covers a wide range of galactocentric distances (with an average position of $\rm 1.5 \,r_e$). In these regards, we find that the radial distance at which the abundance distribution of a galaxy decays the quantity given by the characteristic abundance gradient (0.10 $\rm dex/r_e$) correlate very well with the disc effective radius. Using this new normalisation scale (the abundance scale length, $r_{O/H}$), the distributions of radial positions of the inner drop and outer flattening get significantly narrower, especially in the case of the outer flattening, showing a similar radial position of these features for all the galaxies ($\rm 0.5 \,r_{O/H}$ for the inner drop and $\rm 1.5 \,r_{O/H}$ for the outer flattening). Finally, in agreement with \citet{sanchezmenguiano2016}, we find a dependence of the presence and depth of the inner drop with the integrated stellar mass of the galaxies (with the most massive systems presenting the deepest abundance drop) and an absence of such dependence in the case of the outer flattening.

The high resolution of MUSE data allows for the detection of a large number of \hii\,regions in individual galaxies, providing an accurate derivation of the dispersion around the abundance gradient (as a measurement of the degree of inhomogeneity in the abundance distribution). Thanks to the number of galaxies in our sample, we can perform, for the first time, an extended study analysing possible dependences of the dispersion of the abundance gradient with other properties of the galaxies, such as the integrated stellar mass, the sSFR or the presence of bars. This type of analysis can help us to constraint the origin of the scatter of the abundance gradients. We find no significant dependence of the dispersion with any of the analysed parameters (presence of bars, density of the environment, morphological type, sSFR and stellar mass). Indeed, the derived values are compatible with the uncertainties associated with the derivation of the abundances. This result indicates that statistically, galaxies present a high degree of homogeneity in the abundance distribution, and the possible existing azimuthal variations do not affect significantly the general abundance gradient.

In order to test the robustness of our results, we have reproduced the analysis using two other calibrators: the one described in \citet[][D16]{dopita2016} based on the use of H$\alpha$, [\nii] and [\sii] emission lines; and the calibration proposed in \citet[][]{marino2013} for the N2 index (M13-N2). In general, we obtained a very good agreement between the results based on the three calibrators. The abundance distribution in all cases presents a characteristic slope very similar for all spiral galaxies, although the actual value changes ($\rm -0.20 \,dex/r_e$ for D16 and $\rm -0.04 \,dex/r_e$ for M13-N2). In addition, the stacking of the abundance distribution when normalising to the abundance scalelength $r_{\rm O/H}$ presents the same shape for all calibrators, deriving very similar values for the position of the inner drop and the outer flattening. The most significant difference lies on the number of inner drops and outer flattenings detected in the abundance distribution, that is reduced when using D16 and M13-N2. At this point we should bear in mind that D16 and M13-N2 are indirect calibrators not using any oxygen emission line. The main calibrator used in this work (O3N2) is the only one using such information available for the MUSE wavelength range. For this reason we find our results more reliable than those found using the other calibrators. However, the overall distributions are rather similar in the three cases, strengthening our conclusions.

The results presented in this paper show the potential of high resolution data in statistically significant samples to characterise the abundance distribution of spiral galaxies, proving more accurate constraints to chemical evolution models to explain the observed behaviour in our attempt to understand how discs form.

\vspace{0.5cm}
\begin{acknowledgements}
This study is based on observations made with ESO Telescopes at the Paranal Observatory (programmes 60.A-9329(A), 095.D-0172(A), 95.D-0091(A), 95.D-0091(B), 096.D-0263(A), 96.D-0296(A), 97.D-0408(A) and 98.D-0115(A)) and has also made use of the services of the ESO Science Archive Facility (programmes 60.A-9319(A), 60.A-9100(B), 60.A-9329(A), 60.A-9339(A), 60.A-9301(A), 196.B-0578(A) and 094.B-0733(B)).\\

We would like to thank the anonymous referee for comments which helped to improve the content of the paper.\\

We acknowledge financial support from the Spanish {\em Ministerio de Econom\'ia y Competitividad (MINECO)} via grants AYA2012-31935, and from the `Junta de Andaluc\'ia' local government through the FQM-108 project. We also acknowledge support to the ConaCyt funding programme 180125 and DGAPA IA100815. L.G. is supported in part by the US National Science Foundation under Grant AST-1311862.  T.K. acknowledges support through the Sofja Kovalevskaja Award to P. Schady from the Alexander von Humboldt Foundation of Germany.\\ 

We acknowledge the usage of the HyperLeda database (http://leda.univ-lyon1.fr). This research also makes use of python (\url{http://www.python.org}), of Matplotlib \citep[][]{hunter2007}, a suite of open-source python modules that provides a framework for creating scientific plots, and Astropy, a community-developed core Python package for Astronomy \citep[][]{astropy2013}.

\end{acknowledgements}

\bibliographystyle{aa} 
\bibliography{bibliography}

\makeatletter\renewcommand\theenumi{\@alph\c@enumi}\makeatother
\renewcommand\labelenumi{(\theenumi)}

\clearpage
\onecolumn

\appendix
\section{Sample characterisation (I). General properties.}\label{sec:appendix1} 

In this section we present a table with the main properties of the galaxies analysed in this work. From left to right the columns correspond to

\vspace{0.5cm}
(a) the galaxy name; \par\vspace{0.1cm}
(b)$-$(c) right ascension (in hours) and declination (in degrees) referred to J2000; \par\vspace{0.1cm}
(d) morphological type from {\it HyperLeda}; \par\vspace{0.1cm}
(e) redshift; \par\vspace{0.1cm}
(f) disc effective radius ($r_e$, in kpc), derived from the surface brightness profile of the galaxies by fitting ellipses to the $g$-band light distribution (see Sect.~\ref{sec:analysis3} for details); \par\vspace{0.1cm}
(g)$-$(h) position and inclination angles of the galaxy discs (in degrees), derived by fitting ellipses to the $g$-band image (details are given in Sect.~\ref{sec:analysis3}); \par\vspace{0.1cm}
(i) absolute $r$-band magnitudes in the AB magnitude system derived from the flux density in a $r$-band image recovered from the data; \par\vspace{0.1cm}
(j) $g-r$ colours; \par\vspace{0.1cm}
(k)$-$(l) integrated stellar mass (in $\rm \log M_{\odot}$) and mass density (in $\rm \log M_{\odot}/pc^2$), calculated using the mass-luminosity ratio presented in \citet{bell2001}, see Sect.~\ref{masses} for more information; \par\vspace{0.1cm}
(m) specific star formation rate (in yr$^{-1}$), derived by adopting the classical relation presented in \citet{kennicutt1998} based on dust-corrected H$\alpha$ luminosities (more details are given in Sect.~\ref{sec:scatter}); \par\vspace{0.1cm}
(n) classification according to the density of the environment (isolated, S; part of a group of at least three galaxies, G; in pair, P; and with evidence of real interactions, I) based on information found in the literature (references are given in the notes below).

\vspace{0.5cm}
\noindent
{\bf Notes.} $^{(1)}$ Galaxy discarded from the analysis of the abundance distribution performed from Sect~\ref{sec:slope} on for: (a) having a very powerful AGN in the centre that can contaminate the emission coming from the \hii\,regions, (b) having a ring-like structure, what affects the accuracy of the derivation of the abundance gradient, (c) being very distorted due to a recent interacting, and (d) being in a very advanced stage of merging. $^{(2)}$ Despite this classification by {\it HyperLeda}, it is clear the presence of a spiral structure observed with our data. $^{(3)}$ References: (a) \citet{vorontsov1977}, (b) \citet{tully2015}, (c) \citet{focardi2006}, (d) \citet{mendoza2015}, (e) \citet{mazzuca2006}, (f) \citet{habergham2012}, (g) \citet{white1999}, (h) \citet{ramella2002}, (i) \citet{focardi2002}, (j) \citet{yang2007}, (k) \citet{dopita2002}, (l) \citet{casasola2004}, (m) \citet{martinez2010}, (n) \citet{foster2014}, (o) \citet{keel2013}, (p) \citet{ganda2007}, (q) \citet{vandenbergh2007}, (r) \citet{berlind2006}, (s) \citet{knapen2014}, (t) \citet{hernandez2016}, (u) \citet{madore2009}, (v) \citet{couto2006}, and (w) \citet{merchan2005}.

\begin{center}
\begin{longtab}
\footnotesize
\tabcolsep=0.25cm
\LTcapwidth=\textwidth
\begin{landscape}
\begin{longtable}{l@{\hspace{0.6cm}}cclcccccccccc}
\caption{Fundamental properties of the galaxy sample. Details are given in the main text above.}\\
\hline\hline\\
Name & RA & Dec & Morph & z & $r_e$ & PA & i & $M_{r}$ & $g-r$ & $\log$ Mass & $\log \sigma$ Mass & sSFR & Density of  \\[0.1cm]
  & [hours] & [$\degree$] & type & & [kpc] & [$\degree$] & [$\degree$] & [mag] & [mag] & [M$_{\odot}$] & [M$_{\odot}$/pc$^2$] & [yr$^{-1}$] & environment  \\[0.2cm]
{\tiny(a)} & {\tiny(b)} & {\tiny(c)} & {\tiny(d)} & {\tiny(e)} & {\tiny(f)} & {\tiny(g)} & {\tiny(h)} & {\tiny(i)} & {\tiny(j)} & {\tiny(k)} & {\tiny(l)} & {\tiny(m)} & {\tiny(n)} \\[0.1cm]
\hline\\
\endfirsthead
\caption{Continued.}\\
\hline\hline\\
Name & RA & Dec & Morph & z & $r_e$ & PA & i & $M_{r}$ & $g-r$ & $\log$ Mass & $\log \sigma$ Mass & sSFR & Density of  \\[0.1cm]
  & [hours] & [$\degree$] & type & & [kpc] & [$\degree$] & [$\degree$] & [mag] & [mag] & [M$_{\odot}$] & [M$_{\odot}$/pc$^2$] & [yr$^{-1}$] & environment  \\[0.2cm]
{\tiny(a)} & {\tiny(b)} & {\tiny(c)} & {\tiny(d)} & {\tiny(e)} & {\tiny(f)} & {\tiny(g)} & {\tiny(h)} & {\tiny(i)} & {\tiny(j)} & {\tiny(k)} & {\tiny(l)} & {\tiny(m)} & {\tiny(n)} \\[0.1cm]
\hline\\
\endhead
\hline\\
\endfoot
\hline\\
\endlastfoot
2MASXJ00351050+2315184   & $ 00.59 $ & $ +23.26 $ &  Sc   &  0.037  &  17.2  & $ +81 $ &  63  & $ -21.5 $ & $ -0.44 $ &  10.9  &  1.7  & $ -10.9 $ &  S  \\
2MASXJ01460987-6727579   & $ 01.77 $ & $ -67.47 $ &  SABab   &  0.087  &  10.8  & $ +5 $ &  41  & $ -22.0 $ & $ 0.47 $ &  11.3  &  2.2  & $ -10.8 $ &  S  \\
2MASXJ01504127-1431032   & $ 01.84 $ & $ -14.52 $ &  ---   &  0.034  &  6.2  & $ -72 $ &  57  & $ -21.0 $ & $ 0.52 $ &  10.8  &  2.4  & $ -10.3 $ &  S  \\
2MASXJ06380745-7543288   & $ 06.64 $ & $ -75.72 $ &  ---   &  0.040  &  6.3  & $ +65 $ &  35  & $ -21.4 $ & $ 0.51 $ &  11.0  &  2.4  & $ -10.4 $ &  S  \\
2MASXJ07353554-6246099   & $ 07.59 $ & $ -62.77 $ &  Scd   &  0.011  &  1.3  & $ -60 $ &  39  & $ -18.8 $ & $ 0.70 $ &  9.8  &  2.6  & $ -10.3 $ &  S  \\
2MASXJ09092454-0443026   & $ 09.16 $ & $ -04.72 $ &  SBa   &  0.046  &  4.8  & $ -89 $ &  24  & $ -21.1 $ & $ 0.26 $ &  10.6  &  2.2  & $ -10.3 $ &  S  \\
2MASXJ09151727-2536001   & $ 09.25 $ & $ -25.60 $ &  SBa   &  0.053  &  5.8  & $ -85 $ &  51  & $ -21.2 $ & $ 0.46 $ &  10.9  &  2.5  & $ -10.1 $ &  S  \\
2MASXJ10064866-0741124   & $ 10.11 $ & $ -07.69 $ &  Sbc   &  0.055  &  14.0  & $ -21 $ &  59  & $ -22.1 $ & $ 0.42 $ &  11.1  &  2.0  & $ -10.6 $ &  S  \\
2MASXJ23451480-2947009   & $ 23.75 $ & $ -29.78 $ &  ---   &  0.034  &  5.0  & $ -59 $ &  61  & $ -20.2 $ & $ 0.43 $ &  10.6  &  2.4  & $ -10.1 $ &  S  \\
APMUKS(BJ)B033105.19-262232.9   & $ 03.55 $ & $ -26.21 $ &  ---   &  0.040  &  4.6  & $ +9 $ &  39  & $ -19.6 $ & $ 0.36 $ &  10.3  &  2.0  & $ -10.3 $ &  S  \\
CGCG023-005   & $ 15.96 $ & $ +01.11 $ &  ---   &  0.033  &  5.6  & $ +15 $ &  52  & $ -21.2 $ & $ 0.32 $ &  10.9  &  2.5  & $ -10.2 $ &  S  \\
CGCG044-035   & $ 13.23 $ & $ +06.96 $ &  SBbc   &  0.024  &  6.0  & $ -32 $ &  68  & $ -20.1 $ & $ 0.37 $ &  10.5  &  2.3  & $ -10.4 $ &  G $\rm^{(3h)}$ \\
CGCG048-051   & $ 14.89 $ & $ +02.97 $ &  Sm   &  0.028  &  5.7  & $ +0 $ &  50  & $ -20.9 $ & $ 0.46 $ &  10.8  &  2.3  & $ -10.7 $ &  S  \\
CGCG063-098   & $ 09.90 $ & $ +09.19 $ &  Sbc   &  0.030  &  3.5  & $ +84 $ &  28  & $ -21.2 $ & $ 0.60 $ &  10.8  &  2.7  & $ -11.2 $ &  S  \\
ESO075-G049   & $ 22.03 $ & $ -70.04 $ &  ---   &  0.038  &  5.0  & $ +80 $ &  38  & $ -21.8 $ & $ 0.44 $ &  11.1  &  2.7  & $ -10.3 $ &  S  \\
ESO184-G082   & $ 19.58 $ & $ -52.84 $ &  SBd   &  0.009  &  2.0  & $ -35 $ &  30  & $ -18.6 $ & $ 0.21 $ &  9.8  &  2.2  & $ -10.1 $ &  S  \\
ESO235-G37   & $ 21.03 $ & $ -48.27 $ &  Sbc   &  0.019  &  6.5  & $ -29 $ &  62  & $ -20.3 $ & $ 0.48 $ &  10.5  &  2.1  & $ -10.7 $ &  S  \\
ESO287-G040   & $ 21.62 $ & $ -47.04 $ &  SBbc   &  0.031  &  13.6  & $ -38 $ &  42  & $ -21.1 $ & $ 0.54 $ &  10.9  &  1.7  & $ -10.4 $ &  I $\rm^{(3l,u)}$ \\
ESO298-G28   & $ 02.33 $ & $ -37.82 $ &  Sc   &  0.017  &  12.6  & $ +21 $ &  59  & $ -21.9 $ & $ 0.45 $ &  11.2  &  2.1  & $ -10.8 $ &  S  \\
ESO462-G009   & $ 20.36 $ & $ -31.29 $ &  Sb   &  0.019  &  5.0  & $ -70 $ &  40  & $ -20.9 $ & $ 0.51 $ &  10.7  &  2.3  & $ -10.4 $ &  S  \\
ESO478-G006   & $ 02.16 $ & $ -23.42 $ &  SBc   &  0.018  &  7.8  & $ -75 $ &  52  & $ -21.6 $ & $ 0.45 $ &  11.3  &  2.6  & $ -10.1 $ &  S  \\
ESO488-G30   & $ 05.81 $ & $ -24.38 $ &  SBa   &  0.040  &  16.0  & $ +52 $ &  37  & $ -24.0 $ & $ 0.22 $ &  11.6  &  2.2  & $ -11.0 $ &  G $\rm^{(3b)}$ \\
ESO506-G004   & $ 12.36 $ & $ -24.17 $ &  Sc   &  0.013  &  5.7  & $ +85 $ &  60  & $ -21.2 $ & $ 0.61 $ &  11.0  &  2.7  & $ -10.4 $ &  S  \\
ESO510-G034   & $ 14.00 $ & $ -25.38 $ &  ---   &  0.038  &  9.5  & $ -75 $ &  66  & $ -21.9 $ & $ 0.59 $ &  11.0  &  2.3  & $ -10.9 $ &  S  \\
ESO570-G020   & $ 11.39 $ & $ -22.27 $ &  SBcd   &  0.028  &  6.8  & $ +71 $ &  51  & $ -21.2 $ & $ 0.51 $ &  11.0  &  2.4  & $ -10.4 $ &  S  \\
ESO602-G025   & $ 22.52 $ & $ -19.03 $ &  Scd   &  0.025  &  15.1  & $ -10 $ &  69  & $ -22.1 $ & $ 0.38 $ &  11.3  &  2.3  & $ -10.2 $ &  S  \\
GALEXASCJ125413.02-073849.5   & $ 12.90 $ & $ -07.65 $ &  Sc   &  0.014  &  4.4  & $ +53 $ &  28  & $ -18.4 $ & $ 0.28 $ &  10.0  &  1.6  & $ -10.4 $ &  S  \\
GALEXASCJ195932.30-565932.4   & $ 19.99 $ & $ -56.99 $ &  ---   &  0.057  &  9.9  & $ +60 $ &  41  & $ -22.0 $ & $ 0.32 $ &  11.2  &  2.2  & $ -10.4 $ &  S  \\
IC0344   & $ 03.69 $ & $ -04.67 $ &  SABcd   &  0.018  &  6.4  & $ +45 $ &  59  & $ -20.9 $ & $ 0.32 $ &  10.8  &  2.4  & $ -10.3 $ &  G $\rm^{(3h)}$ \\
IC0577   & $ 09.93 $ & $ +10.50 $ &  SABd   &  0.030  &  7.0  & $ -5 $ &  27  & $ -22.5 $ & $ 0.20 $ &  11.3  &  2.6  & $ -10.4 $ &  P $\rm^{(3b,t)}$ \\
IC1158   & $ 16.03 $ & $ +01.71 $ &  SABd   &  0.006  &  2.3  & $ -47 $ &  55  & $ -19.4 $ & $ 0.42 $ &  10.0  &  2.5  & $ -10.6 $ &  S  \\
IC2160   & $ 05.92 $ & $ -76.92 $ &  Sc   &  0.016  &  5.2  & $ -75 $ &  39  & $ -21.5 $ & $ 0.42 $ &  11.0  &  2.6  & $ -10.3 $ &  S  \\
IC5179   & $ 22.27 $ & $ -36.84 $ &  SBbc   &  0.011  &  7.6  & $ +55 $ &  61  & $ -21.9 $ & $ 0.48 $ &  11.4  &  2.8  & $ -9.9 $ &  S  \\
LCRSB232926.0-413242   & $ 23.54 $ & $ -41.27 $ &  ---   &  0.049  &  10.4  & $ -37 $ &  63  & $ -21.1 $ & $ 0.34 $ &  10.8  &  2.0  & $ -10.5 $ &  S  \\
LCSBS0801P   & $ 04.92 $ & $ -20.00 $ &  ---   &  0.080  &  13.8  & $ +30 $ &  38  & $ -22.2 $ & $ 0.44 $ &  11.3  &  2.0  & $ -10.5 $ &  S  \\
MCG+00-06-003   & $ 01.95 $ & $ -00.21 $ &  ---   &  0.046  &  8.0  & $ +36 $ &  61  & $ -21.6 $ & $ 0.44 $ &  11.0  &  2.4  & $ -11.0 $ &  S  \\
MCG+03-31-093   & $ 12.29 $ & $ +16.37 $ &  ---   &  0.023  &  4.7  & $ +50 $ &  34  & $ -20.6 $ & $ 0.20 $ &  10.6  &  2.2  & $ -10.4 $ &  S  \\
MCG+04-55-47   & $ 23.71 $ & $ +27.09 $ &  Scd   &  0.025  &  4.7  & $ +7 $ &  49  & $ -20.5 $ & $ 0.42 $ &  10.6  &  2.3  & $ -10.3 $ &  G $\rm^{(3b)}$ \\
MCG-01-33-034   & $ 12.88 $ & $ -09.78 $ &  SBa   &  0.009  &  2.4  & $ -66 $ &  51  & $ -19.6 $ & $ 0.47 $ &  10.0  &  2.3  & $ -10.0 $ &  S  \\
MCG-01-57-021   & $ 22.67 $ & $ -02.42 $ &  Sa   &  0.010  &  8.2  & $ -88 $ &  52  & $ -20.3 $ & $ 0.33 $ &  10.7  &  2.0  & $ -10.6 $ &  S  \\
MCG-02-01-014   & $ 00.07 $ & $ -11.17 $ &  SB0a   &  0.038  &  15.1  & $ -50 $ &  43  & $ -20.9 $ & $ 0.33 $ &  10.9  &  1.6  & $ -10.2 $ &  P $\rm^{(3o)}$ \\
MCG-02-07-33   & $ 02.57 $ & $ -10.84 $ &  ---   &  0.016  &  7.4  & $ +73 $ &  72  & $ -21.4 $ & $ 0.30 $ &  11.0  &  2.6  & $ -10.3 $ &  G $\rm^{(3b)}$ \\
MCG-03-07-040   & $ 02.60 $ & $ -17.26 $ &  Scd   &  0.032  &  9.5  & $ -5 $ &  52  & $ -21.2 $ & $ 0.27 $ &  10.9  &  2.1  & $ -10.2 $ &  P  \\
MCG-03-10-052   & $ 03.91 $ & $ -19.19 $ &  SBa   &  0.025  &  5.3  & $ +28 $ &  55  & $ -21.5 $ & $ 0.50 $ &  11.1  &  2.7  & $ -10.4 $ &  S  \\
NGC0232 $\rm^{(1a)}$  & $ 00.71 $ & $ -23.56 $ &  S0$^0$ $^{(2)}$  &  0.022  &  5.1  & $ +38 $ &  41  & $ -21.3 $ & $ 0.67 $ &  ---  &  ---  &  ---  &  I $\rm^{(3k,l)}$ \\
NGC1080   & $ 02.75 $ & $ -04.71 $ &  Sb   &  0.026  &  9.1  & $ +5 $ &  36  & $ -22.0 $ & $ 0.37 $ &  11.3  &  2.3  & $ -10.3 $ &  S  \\
NGC1309   & $ 03.37 $ & $ -15.40 $ &  Sb   &  0.007  &  3.8  & $ +72 $ &  27  & $ -21.0 $ & $ 0.34 $ &  11.0  &  2.8  & $ -10.2 $ &  S  \\
NGC1516A $\rm^{(1d)}$  & $ 04.14 $ & $ -08.83 $ &  ---   &  0.033  &  6.3  & $ -30 $ &  30  & $ -22.6 $ & $ 0.44 $ &  ---  &  ---  &  ---  &  I $\rm^{(3b)}$ \\
NGC1566   & $ 04.33 $ & $ -54.94 $ &  SBc   &  0.005  &  1.7  & $ +21 $ &  56  & $ -18.6 $ & $ 0.51 $ &  10.1  &  2.8  & $ -10.6 $ &  G $\rm^{(3b)}$ \\
NGC1578   & $ 04.40 $ & $ -51.60 $ &  SBc   &  0.020  &  4.9  & $ -28 $ &  27  & $ -21.9 $ & $ 0.49 $ &  11.1  &  2.6  & $ -10.5 $ &  S  \\
NGC1590   & $ 04.52 $ & $ +07.63 $ &  Sbc   &  0.014  &  3.3  & $ -68 $ &  44  & $ -21.3 $ & $ 0.10 $ &  11.1  &  3.1  & $ -10.2 $ &  G $\rm^{(3b)}$ \\
NGC1591   & $ 04.49 $ & $ -26.71 $ &  SBbc   &  0.014  &  6.2  & $ +33 $ &  50  & $ -21.3 $ & $ 0.17 $ &  11.1  &  2.6  & $ -10.3 $ &  G $\rm^{(3b)}$ \\
NGC1762   & $ 05.06 $ & $ +01.57 $ &  ---   &  0.016  &  7.3  & $ -3 $ &  51  & $ -22.4 $ & $ 0.33 $ &  11.4  &  2.7  & $ -10.5 $ &  S  \\
NGC1954   & $ 05.55 $ & $ -14.06 $ &  SABa   &  0.010  &  6.2  & $ -33 $ &  59  & $ -21.9 $ & $ 0.41 $ &  10.7  &  2.2  & $ -10.2 $ &  G $\rm^{(3b)}$ \\
NGC2370   & $ 07.42 $ & $ +23.78 $ &  Sb   &  0.018  &  7.9  & $ +43 $ &  61  & $ -22.1 $ & $ 0.56 $ &  11.2  &  2.6  & $ -10.3 $ &  G $\rm^{(3b)}$ \\
NGC2466   & $ 07.75 $ & $ -71.41 $ &  Sc   &  0.018  &  5.8  & $ -7 $ &  23  & $ -21.9 $ & $ 0.34 $ &  11.2  &  2.6  & $ -10.0 $ &  S  \\
NGC2906   & $ 09.54 $ & $ +08.44 $ &  Sbc   &  0.007  &  5.0  & $ +82 $ &  56  & $ -21.1 $ & $ 0.50 $ &  10.9  &  2.7  & $ -10.8 $ &  S  \\
NGC2930   & $ 09.63 $ & $ +23.20 $ &  SABbc   &  0.022  &  10.6  & $ -47 $ &  51  & $ -20.9 $ & $ 0.22 $ &  10.5  &  1.5  & $ -9.9 $ &  G $\rm^{(3g,h)}$ \\
NGC3120   & $ 10.09 $ & $ -34.22 $ &  Sc   &  0.009  &  3.9  & $ +0 $ &  52  & $ -20.5 $ & $ 0.45 $ &  10.7  &  2.6  & $ -10.5 $ &  G $\rm^{(3b)}$ \\
NGC3244   & $ 10.42 $ & $ -39.83 $ &  Sbc   &  0.009  &  4.2  & $ +0 $ &  31  & $ -20.0 $ & $ 0.58 $ &  10.7  &  2.4  & $ -10.2 $ &  G $\rm^{(3b)}$ \\
NGC3278   & $ 10.53 $ & $ -39.95 $ &  Sb   &  0.010  &  3.4  & $ +56 $ &  53  & $ -21.1 $ & $ 0.42 $ &  10.9  &  3.0  & $ -10.0 $ &  G $\rm^{(3b)}$ \\
NGC3318   & $ 10.62 $ & $ -41.63 $ &  SB0a   &  0.009  &  7.4  & $ +81 $ &  63  & $ -20.8 $ & $ 0.38 $ &  11.1  &  2.6  & $ -10.1 $ &  S  \\
NGC3363   & $ 10.75 $ & $ +22.08 $ &  SABab   &  0.019  &  6.6  & $ +3 $ &  58  & $ -21.5 $ & $ 0.46 $ &  11.1  &  2.6  & $ -10.5 $ &  S  \\
NGC3389   & $ 10.81 $ & $ +12.53 $ &  Sc   &  0.004  &  5.0  & $ -71 $ &  62  & $ -20.0 $ & $ 0.27 $ &  10.6  &  2.4  & $ -10.2 $ &  G $\rm^{(3b,q)}$ \\
NGC3447 $\rm^{(1c)}$  & $ 10.89 $ & $ +16.77 $ &  S0a   &  0.004  &  2.0  & $ +1 $ &  57  & $ -17.4 $ & $ 0.33 $ &  ---  &  ---  &  ---  &  I $\rm^{(3s,t)}$ \\
NGC3451   & $ 10.91 $ & $ +27.24 $ &  SB0a   &  0.004  &  4.2  & $ +54 $ &  67  & $ -19.9 $ & $ 0.35 $ &  10.5  &  2.5  & $ -10.1 $ &  G $\rm^{(3b)}$ \\
NGC3464   & $ 10.91 $ & $ -21.07 $ &  Sbc   &  0.012  &  9.5  & $ -72 $ &  49  & $ -21.3 $ & $ 0.40 $ &  11.1  &  2.2  & $ -10.4 $ &  S  \\
NGC3512   & $ 11.07 $ & $ +28.04 $ &  SBbc   &  0.005  &  2.6  & $ -50 $ &  33  & $ -19.9 $ & $ 0.49 $ &  10.4  &  2.5  & $ -10.4 $ &  G $\rm^{(3b,f)}$ \\
NGC3905   & $ 11.82 $ & $ -09.73 $ &  SBab   &  0.019  &  11.8  & $ -80 $ &  42  & $ -22.7 $ & $ 0.43 $ &  11.6  &  2.4  & $ -10.5 $ &  S  \\
NGC4451   & $ 12.48 $ & $ +09.26 $ &  SBcd   &  0.003  &  2.1  & $ -13 $ &  54  & $ -19.6 $ & $ 0.37 $ &  10.4  &  2.9  & $ -10.4 $ &  G $\rm^{(3b)}$ \\
NGC4487   & $ 12.52 $ & $ -08.05 $ &  SABc   &  0.003  &  8.0  & $ +70 $ &  54  & $ -20.8 $ & $ 0.38 $ &  10.8  &  2.1  & $ -10.5 $ &  P $\rm^{(3b,p)}$ \\
NGC4651   & $ 12.73 $ & $ +16.39 $ &  ---   &  0.003  &  3.6  & $ +70 $ &  50  & $ -21.5 $ & $ 0.42 $ &  11.0  &  3.0  & $ -10.6 $ &  I $\rm^{(3m,n)}$ \\
NGC4806   & $ 12.94 $ & $ -29.50 $ &  SB0a   &  0.008  &  2.9  & $ +28 $ &  37  & $ -20.0 $ & $ 0.29 $ &  10.6  &  2.6  & $ -10.2 $ &  S  \\
NGC4900   & $ 13.01 $ & $ +02.50 $ &  SBbc   &  0.003  &  7.8  & $ -80 $ &  20  & $ -21.4 $ & $ 0.35 $ &  10.7  &  1.8  & $ -10.3 $ &  G $\rm^{(3b)}$ \\
NGC4965   & $ 13.12 $ & $ -28.23 $ &  SBc   &  0.008  &  13.4  & $ -27 $ &  37  & $ -20.4 $ & $ 0.41 $ &  11.3  &  2.0  & $ -10.4 $ &  P $\rm^{(3b,f)}$ \\
NGC5339   & $ 13.90 $ & $ -07.93 $ &  SBm   &  0.009  &  4.4  & $ +25 $ &  34  & $ -20.8 $ & $ 0.44 $ &  10.7  &  2.4  & $ -10.7 $ &  S  \\
NGC5584   & $ 14.37 $ & $ -00.39 $ &  SABc   &  0.005  &  8.4  & $ -23 $ &  45  & $ -19.0 $ & $ 0.41 $ &  ---  &  ---  & $ -10.3 $ &  S  \\
NGC6708   & $ 18.93 $ & $ -53.72 $ &  SABbc   &  0.009  &  2.1  & $ -6 $ &  41  & $ -21.2 $ & $ 0.38 $ &  10.9  &  3.3  & $ -10.0 $ &  G $\rm^{(3b)}$ \\
NGC6754   & $ 19.19 $ & $ -50.64 $ &  SABbc   &  0.011  &  8.7  & $ +86 $ &  63  & $ -20.9 $ & $ -0.14 $ &  11.0  &  2.4  & $ -10.8 $ &  S  \\
NGC7119A   & $ 21.77 $ & $ -46.52 $ &  SABbc   &  0.032  &  9.9  & $ -45 $ &  57  & $ -22.6 $ & $ 0.39 $ &  11.6  &  2.7  & $ -10.2 $ &  I $\rm^{(3l,v)}$ \\
NGC7469 $\rm^{(1a)}$  & $ 23.05 $ & $ +08.87 $ &  SBb   &  0.016  &  3.4  & $ -59 $ &  38  & $ -22.2 $ & $ 0.43 $ &  ---  &  ---  &  ---  &  I $\rm^{(3b,d)}$ \\
NGC7580 $\rm^{(1c)}$  & $ 23.29 $ & $ +14.00 $ &  S0a   &  0.016  &  2.9  & $ +52 $ &  33  & $ -20.8 $ & $ 0.39 $ &  ---  &  ---  &  ---  &  G $\rm^{(3b,j)}$ \\
NGC7742   & $ 23.74 $ & $ +10.77 $ &  Sb   &  0.006  &  0.9  & $ +0 $ &  0  & $ -20.1 $ & $ 0.50 $ &  10.4  &  3.4  & $ -10.3 $ &  P $\rm^{(3b,e)}$ \\
PGC004701   & $ 01.31 $ & $ -07.45 $ &  Sbc   &  0.018  &  8.5  & $ -18 $ &  28  & $ -20.2 $ & $ 0.41 $ &  10.7  &  1.8  & $ -10.4 $ &  P $\rm^{(3a)}$ \\
PGC055442   & $ 15.56 $ & $ +21.14 $ &  E $^{(2)}$  &  0.024  &  3.8  & $ +15 $ &  46  & $ -21.5 $ & $ 0.55 $ &  10.9  &  2.8  & $ -11.0 $ &  P $\rm^{(3c)}$ \\
PGC128348   & $ 23.55 $ & $ -60.57 $ &  ---   &  0.015  &  4.1  & $ +35 $ &  34  & $ -20.0 $ & $ 0.34 $ &  10.4  &  2.2  & $ -10.4 $ &  S  \\
PGC1285465   & $ 11.00 $ & $ +05.56 $ &  Sc   &  0.056  &  10.6  & $ +10 $ &  59  & $ -21.0 $ & $ 0.49 $ &  10.6  &  1.8  & $ -10.4 $ &  S  \\
SDSSJ011557.69+004135.9   & $ 01.27 $ & $ +00.69 $ &  Sbc   &  0.043  &  6.2  & $ -80 $ &  20  & $ -20.1 $ & $ 0.32 $ &  10.4  &  1.7  & $ -10.4 $ &  G $\rm^{(3r,w)}$ \\
SDSSJ090202.19+101759.7   & $ 09.03 $ & $ +10.30 $ &  ---   &  0.042  &  10.9  & $ +23 $ &  46  & $ -20.3 $ & $ -0.18 $ &  10.3  &  1.2  & $ -10.7 $ &  S  \\
UGC00272   & $ 00.46 $ & $ -01.20 $ &  Sdm   &  0.013  &  8.6  & $ -53 $ &  70  & $ -19.6 $ & $ 0.28 $ &  10.4  &  1.9  & $ -10.4 $ &  G $\rm^{(3h)}$ \\
UGC01333   & $ 01.86 $ & $ +00.26 $ &  Sc   &  0.059  &  19.0  & $ -80 $ &  46  & $ -22.8 $ & $ 0.89 $ &  12.6  &  3.1  & $ -9.7 $ &  S  \\
UGC01395   & $ 01.92 $ & $ +06.61 $ &  Sb   &  0.017  &  7.0  & $ -34 $ &  44  & $ -21.3 $ & $ 0.49 $ &  11.0  &  2.4  & $ -10.5 $ &  S  \\
UGC02019   & $ 02.54 $ & $ +00.62 $ &  Sbc   &  0.021  &  4.5  & $ +67 $ &  38  & $ -21.3 $ & $ 0.17 $ &  11.0  &  2.7  & $ -10.5 $ &  G $\rm^{(3b,r)}$ \\
UGC03634 $\rm^{(1b)}$  & $ 07.03 $ & $ +14.14 $ &  Sbc   &  0.026  &  7.4  & $ -57 $ &  52  & $ -22.3 $ & $ 0.68 $ &  ---  &  ---  &  ---  &  S  \\
UGC05378   & $ 10.01 $ & $ +04.41 $ &  Sb   &  0.014  &  2.9  & $ -79 $ &  59  & $ -19.3 $ & $ 0.30 $ &  10.2  &  2.4  & $ -10.3 $ &  S  \\
UGC05691   & $ 10.49 $ & $ +22.01 $ &  SABb   &  0.054  &  20.0  & $ -49 $ &  44  & $ -22.6 $ & $ 0.48 $ &  11.4  &  1.8  & $ -10.6 $ &  S  \\
UGC06332 $\rm^{(1b)}$  & $ 11.32 $ & $ +20.81 $ &  S0a   &  0.021  &  8.6  & $ -35 $ &  33  & $ -21.8 $ & $ 0.73 $ &  ---  &  ---  &  ---  &  S  \\
UGC09530   & $ 14.80 $ & $ +09.66 $ &  Scd   &  0.029  &  16.8  & $ +89 $ &  46  & $ -22.9 $ & $ -0.09 $ &  11.7  &  2.3  & $ -11.0 $ &  G $\rm^{(3g,i,j)}$ \\
UGC11001   & $ 17.84 $ & $ +14.29 $ &  Sc   &  0.014  &  6.8  & $ -48 $ &  67  & $ -21.0 $ & $ 0.36 $ &  11.0  &  2.6  & $ -10.3 $ &  G $\rm^{(3b)}$ \\
UGC11214   & $ 18.38 $ & $ +12.43 $ &  Sbc   &  0.009  &  4.8  & $ +0 $ &  20  & $ -20.5 $ & $ 0.34 $ &  10.6  &  2.2  & $ -10.4 $ &  S  \\
UGC11816   & $ 21.82 $ & $ +00.45 $ &  Sb   &  0.016  &  6.2  & $ +0 $ &  18  & $ -20.8 $ & $ 0.25 $ &  10.9  &  2.2  & $ -10.6 $ &  S  \\
UGC12158   & $ 22.70 $ & $ +20.00 $ &  SABc   &  0.031  &  11.0  & $ +0 $ &  0  & $ -21.9 $ & $ 0.42 $ &  11.2  &  2.0  & $ -10.9 $ &  S  \\
\label{table1}
\end{longtable}
\end{landscape}
\end{longtab}
\end{center}
\newpage

\section{Sample characterisation (II). Oxygen abundance information.}\label{sec:appendix2} 

In this section we present a table with information derived in this work related to the oxygen abundance radial distribution for the galaxies in the sample. From left to right the columns correspond to

\vspace{0.5cm}
(a) the galaxy name; \par\vspace{0.1cm}
(b) slope of the oxygen abundance gradient; \par\vspace{0.1cm}
(c) zero-point of the oxygen abundance gradient; \par\vspace{0.1cm}
(d) galactocentric distance at which the inner drop in the abundances appear ($h_D$), in units of $r_e$ (if this feature is present); \par\vspace{0.1cm}
(e) galactocentric distance at which the radial gradients flatten ($h_F$), in units of $r_e$ (if this feature is present); \par\vspace{0.1cm}
(f) abundance scale length ($r_{O/H}$), defined as the radial position at which the abundance distribution of a galaxy decays 0.09 dex (given by the decrease in the abundance in one disc effective radius attending to the common abundance gradient found, see Sect.~\ref{masses} for details); \par\vspace{0.1cm}
(g) $h_D$, in units of $r_{O/H}$; \par\vspace{0.1cm}
(h) $h_F$, in units of $r_{O/H}$; \par\vspace{0.1cm}
(i) scatter relative to the oxygen abundance gradient (see Sect.~\ref{sec:scatter} for more information);

\vspace{0.5cm}
\noindent
{\bf Notes.} $\;^\dagger$ galaxies for which the derivation of $r_{O/H}$ is infeasible because the radial abundance distribution displays a very flat profile. $\;^\star$ Galaxy discarded from the analysis of the abundance distribution performed from Sect~\ref{sec:slope} on. See Notes (1) in Table~\ref{table1} (Appendix~\ref{sec:appendix1}) for details.

\vspace{1cm}
\begin{center}
\begin{longtab}
\tabcolsep=0.30cm
\LTcapwidth=\textwidth
\begin{longtable}{lcccccccc}
\caption{Oxygen abundance information of the galaxy sample. Details are given in the main text above.}\\
\hline\hline\\
Name & Slope & Zero-point & $h_D$ & $h_F$ & $R_{O/H}$ & $h_D$ & $h_F$ & Scatter \\[0.1cm]
  & [dex/r$\rm _e$] & [dex] & [r$\rm _e$] & [r$\rm _e$] & [kpc] & [r$\rm _{O/H}$] & [r$\rm _{O/H}$] & [dex] \\[0.2cm]
{\tiny(a)} & {\tiny(b)} & {\tiny(c)} & {\tiny(d)} & {\tiny(e)} & {\tiny(f)} & {\tiny(g)} & {\tiny(h)} & {\tiny(i)}\\[0.1cm]
\hline\\
\endfirsthead
\caption{Continued.}\\
\hline\hline\\
Name & Slope & Zero-point & $h_D$ & $h_F$ & $R_{O/H}$ & $h_D$ & $h_F$ & Scatter \\[0.1cm]
  & [dex/r$\rm _e$] & [dex] & [r$\rm _e$] & [r$\rm _e$] & [kpc] & [r$\rm _{O/H}$] & [r$\rm _{O/H}$] & [dex] \\[0.2cm]
{\tiny(a)} & {\tiny(b)} & {\tiny(c)} & {\tiny(d)} & {\tiny(e)} & {\tiny(f)} & {\tiny(g)} & {\tiny(h)} & {\tiny(i)}\\[0.1cm]
\hline\\
\endhead
\hline\\
\endfoot
\hline\\
\endlastfoot
2MASXJ00351050+2315184  & $ -0.27 $ &  8.56  &  ---  &  ---  &  5.0  &  ---  &  ---  &  0.053 \\
2MASXJ01460987-6727579  & $ -0.08 $ &  8.60  &  ---  &  ---  &  11.1  &  ---  &  ---  &  0.041 \\
2MASXJ01504127-1431032  & $ -0.17 $ &  8.73  &  0.8  &  2.3  &  9.4  &  0.5  &  1.5  &  0.036 \\
2MASXJ06380745-7543288  & $ -0.15 $ &  8.73  &  0.6  &  2.8  &  7.6  &  0.5  &  2.3  &  0.062 \\
2MASXJ07353554-6246099  & $ -0.07 $ &  8.64  &  0.9  &  ---  &  1.5  &  0.7  &  ---  &  0.052 \\
2MASXJ09092454-0443026  & $ -0.13 $ &  8.57  &  ---  &  ---  &  3.9  &  ---  &  ---  &  0.030 \\
2MASXJ09151727-2536001  & $ -0.03 $ &  8.58  &  ---  &  ---  &  10.3  &  ---  &  ---  &  0.024 \\
2MASXJ10064866-0741124  & $ -0.09 $ &  8.54  &  ---  &  1.1  &  16.2  &  ---  &  1.0  &  0.038 \\
2MASXJ23451480-2947009  & $ -0.14 $ &  8.51  &  ---  &  1.3  &  4.2  &  ---  &  1.6  &  0.092 \\
APMUKS(BJ)B033105.19-262232.9  & $ -0.09 $ &  8.62  &  0.5  &  ---  &  6.1  &  0.4  &  ---  &  0.022 \\
CGCG023-005  & $ -0.14 $ &  8.61  &  ---  &  1.4  &  5.4  &  ---  &  1.5  &  0.049 \\
CGCG044-035  & $ -0.12 $ &  8.51  &  ---  &  ---  &  6.1  &  ---  &  ---  &  0.033 \\
CGCG048-051$\;^\dagger$  & $ -0.01 $ &  8.62  &  ---  &  ---  &  ---  &  ---  &  ---  &  0.045 \\
CGCG063-098  & $ -0.07 $ &  8.68  &  ---  &  ---  &  6.0  &  ---  &  ---  &  0.022 \\
ESO075-G049  & $ -0.05 $ &  8.60  &  ---  &  ---  &  11.0  &  ---  &  ---  &  0.036 \\
ESO184-G082$\;^\dagger$  & $ 0.00 $ &  8.34  &  ---  &  ---  &  3.2  &  ---  &  ---  &  0.061 \\
ESO235-G37  & $ -0.09 $ &  8.57  &  ---  &  ---  &  7.8  &  ---  &  ---  &  0.038 \\
ESO287-G040  & $ -0.14 $ &  8.68  &  ---  &  ---  &  17.0  &  ---  &  ---  &  0.062 \\
ESO298-G28  & $ -0.19 $ &  8.74  &  0.8  &  ---  &  15.1  &  0.7  &  ---  &  0.042 \\
ESO462-G009$\;^\dagger$  & $ 0.00 $ &  8.56  &  ---  &  ---  &  ---  &  ---  &  ---  &  0.058 \\
ESO478-G006  & $ -0.18 $ &  8.74  &  0.6  &  1.6  &  10.3  &  0.5  &  1.2  &  0.021 \\
ESO488-G30  & $ -0.15 $ &  8.69  &  ---  &  ---  &  17.5  &  ---  &  ---  &  0.062 \\
ESO506-G004  & $ -0.06 $ &  8.68  &  ---  &  ---  &  11.0  &  ---  &  ---  &  0.046 \\
ESO510-G034  & $ -0.06 $ &  8.62  &  ---  &  ---  &  14.6  &  ---  &  ---  &  0.022 \\
ESO570-G020  & $ -0.16 $ &  8.66  &  0.4  &  ---  &  7.4  &  0.4  &  ---  &  0.043 \\
ESO602-G025  & $ -0.24 $ &  8.66  &  0.4  &  0.7  &  9.9  &  0.7  &  1.0  &  0.026 \\
GALEXASCJ125413.02-073849.5  & $ -0.14 $ &  8.40  &  ---  &  1.0  &  4.0  &  ---  &  1.1  &  0.060 \\
GALEXASCJ195932.30-565932.4  & $ -0.21 $ &  8.81  &  0.6  &  2.0  &  11.8  &  0.5  &  1.7  &  0.066 \\
IC0344  & $ -0.19 $ &  8.54  &  ---  &  0.9  &  5.1  &  ---  &  1.2  &  0.060 \\
IC0577  & $ -0.14 $ &  8.65  &  0.4  &  ---  &  7.4  &  0.4  &  ---  &  0.033 \\
IC1158  & $ -0.04 $ &  8.60  &  ---  &  ---  &  2.8  &  ---  &  ---  &  0.092 \\
IC2160  & $ -0.14 $ &  8.67  &  1.0  &  2.5  &  9.6  &  0.5  &  1.3  &  0.054 \\
IC5179  & $ -0.15 $ &  8.67  &  0.3  &  1.2  &  8.2  &  0.3  &  1.1  &  0.026 \\
LCRSB232926.0-413242  & $ -0.16 $ &  8.59  &  ---  &  2.1  &  6.9  &  ---  &  3.1  &  0.045 \\
LCSBS0801P  & $ -0.20 $ &  8.73  &  ---  &  ---  &  10.9  &  ---  &  ---  &  0.057 \\
MCG-01-33-034$\;^\dagger$  & $ -0.01 $ &  8.40  &  ---  &  ---  &  ---  &  ---  &  ---  &  0.081 \\
MCG-01-57-021  & $ -0.23 $ &  8.63  &  ---  &  ---  &  4.5  &  ---  &  ---  &  0.051 \\
MCG-02-01-014  & $ -0.26 $ &  8.67  &  ---  &  ---  &  9.2  &  ---  &  ---  &  0.064 \\
MCG-02-07-33  & $ -0.08 $ &  8.61  &  ---  &  ---  &  10.1  &  ---  &  ---  &  0.075 \\
MCG-03-07-040  & $ -0.23 $ &  8.64  &  ---  &  1.6  &  4.7  &  ---  &  3.2  &  0.072 \\
MCG-03-10-052  & $ -0.06 $ &  8.56  &  ---  &  ---  &  11.3  &  ---  &  ---  &  0.049 \\
MCG+00-06-003  & $ -0.13 $ &  8.71  &  ---  &  ---  &  16.0  &  ---  &  ---  &  0.049 \\
MCG+03-31-093  & $ -0.11 $ &  8.61  &  ---  &  ---  &  4.4  &  ---  &  ---  &  0.054 \\
MCG+04-55-47  & $ -0.06 $ &  8.61  &  0.4  &  ---  &  8.3  &  0.2  &  ---  &  0.027 \\
NGC0232$\;^{\dagger\,\star}$ & $ +0.01 $ &  8.56  &  ---  &  ---  &  ---  &  ---  &  ---  &  --- \\
NGC1080  & $ -0.12 $ &  8.67  &  0.4  &  ---  &  10.3  &  0.3  &  ---  &  0.048 \\
NGC1309  & $ -0.17 $ &  8.61  &  0.4  &  1.0  &  3.4  &  0.4  &  1.1  &  0.024 \\
NGC1516A$\;^\star$ & $ -0.05 $ &  8.57  &  ---  &  ---  &  5.6  &  ---  &  ---  &  --- \\
NGC1566$\;^\dagger$  & $ 0.00 $ &  8.59  &  ---  &  ---  &  ---  &  ---  &  ---  &  0.033 \\
NGC1578$\;^\dagger$  & $ -0.03 $ &  8.62  &  ---  &  ---  &  ---  &  ---  &  ---  &  0.049 \\
NGC1590$\;^\dagger$  & $ -0.02 $ &  8.62  &  ---  &  ---  &  ---  &  ---  &  ---  &  0.017 \\
NGC1591  & $ -0.12 $ &  8.60  &  ---  &  0.9  &  6.3  &  ---  &  0.9  &  0.047 \\
NGC1762  & $ -0.12 $ &  8.69  &  0.6  &  ---  &  11.1  &  0.4  &  ---  &  0.040 \\
NGC1954  & $ -0.12 $ &  8.73  &  ---  &  2.6  &  10.9  &  ---  &  1.5  &  0.086 \\
NGC2370  & $ -0.09 $ &  8.66  &  0.5  &  ---  &  11.6  &  0.4  &  ---  &  0.043 \\
NGC2466  & $ -0.14 $ &  8.68  &  0.4  &  2.0  &  7.3  &  0.4  &  1.6  &  0.040 \\
NGC2906  & $ -0.04 $ &  8.60  &  ---  &  ---  &  6.2  &  ---  &  ---  &  0.056 \\
NGC2930  & $ -0.28 $ &  8.52  &  ---  &  ---  &  4.0  &  ---  &  ---  &  0.040 \\
NGC3120  & $ -0.10 $ &  8.68  &  0.8  &  ---  &  6.3  &  0.5  &  ---  &  0.062 \\
NGC3244  & $ -0.19 $ &  8.69  &  0.4  &  ---  &  4.0  &  0.4  &  ---  &  0.040 \\
NGC3278  & $ -0.09 $ &  8.64  &  0.4  &  1.2  &  4.6  &  0.3  &  0.9  &  0.015 \\
NGC3318  & $ -0.10 $ &  8.59  &  ---  &  ---  &  8.4  &  ---  &  ---  &  0.041 \\
NGC3363  & $ -0.07 $ &  8.68  &  0.9  &  ---  &  13.6  &  0.4  &  ---  &  0.049 \\
NGC3389  & $ -0.29 $ &  8.58  &  ---  &  0.7  &  3.0  &  ---  &  1.2  &  0.101 \\
NGC3447$\;^\star$ & $ -0.03 $ &  8.42  &  ---  &  ---  &  1.5  &  ---  &  ---  &  --- \\
NGC3451  & $ -0.03 $ &  8.51  &  ---  &  ---  &  4.8  &  ---  &  ---  &  0.049 \\
NGC3464  & $ -0.23 $ &  8.75  &  0.6  &  ---  &  9.1  &  0.6  &  ---  &  0.050 \\
NGC3512  & $ -0.09 $ &  8.66  &  0.6  &  ---  &  3.5  &  0.5  &  ---  &  0.049 \\
NGC3905  & $ -0.22 $ &  8.80  &  0.8  &  ---  &  14.7  &  0.6  &  ---  &  0.057 \\
NGC4451  & $ -0.13 $ &  8.66  &  0.4  &  1.2  &  2.2  &  0.4  &  1.1  &  0.025 \\
NGC4487  & $ -0.09 $ &  8.55  &  ---  &  ---  &  6.9  &  ---  &  ---  &  0.062 \\
NGC4651  & $ -0.17 $ &  8.72  &  0.8  &  1.4  &  4.4  &  0.7  &  1.1  &  0.047 \\
NGC4806  & $ -0.08 $ &  8.55  &  ---  &  ---  &  3.3  &  ---  &  ---  &  0.053 \\
NGC4900  & $ -0.30 $ &  8.65  &  0.3  &  ---  &  3.8  &  0.6  &  ---  &  0.046 \\
NGC4965  & $ -0.10 $ &  8.50  &  ---  &  ---  &  10.0  &  ---  &  ---  &  0.069 \\
NGC5339  & $ -0.06 $ &  8.63  &  ---  &  ---  &  6.5  &  ---  &  ---  &  0.031 \\
NGC5584  & $ -0.36 $ &  8.60  &  0.1  &  0.3  &  2.6  &  0.2  &  1.1  &  0.036 \\
NGC6708  & $ -0.14 $ &  8.67  &  0.5  &  1.6  &  2.5  &  0.4  &  1.3  &  0.036 \\
NGC6754  & $ -0.12 $ &  8.67  &  0.4  &  ---  &  10.3  &  0.3  &  ---  &  0.049 \\
NGC7119A  & $ -0.15 $ &  8.66  &  0.5  &  1.6  &  11.9  &  0.4  &  1.3  &  0.031 \\
NGC7469$\;^{\dagger\,\star}$ & $ 0.00 $ &  8.53  &  ---  &  ---  &  ---  &  ---  &  ---  &  --- \\
NGC7580$\;^\star$ & $ -0.03 $ &  8.57  &  ---  &  ---  &  2.1  &  ---  &  ---  &  --- \\
NGC7742  & $ -0.04 $ &  8.60  &  ---  &  ---  &  2.5  &  ---  &  ---  &  0.040 \\
PGC004701  & $ -0.11 $ &  8.57  &  ---  &  ---  &  7.3  &  ---  &  ---  &  0.052 \\
PGC055442  & $ -0.05 $ &  8.56  &  ---  &  ---  &  7.0  &  ---  &  ---  &  0.058 \\
PGC128348  & $ -0.09 $ &  8.54  &  ---  &  ---  &  5.5  &  ---  &  ---  &  0.052 \\
PGC1285465  & $ -0.15 $ &  8.62  &  ---  &  ---  &  10.4  &  ---  &  ---  &  0.031 \\
SDSSJ011557.69+004135.9  & $ -0.07 $ &  8.43  &  ---  &  ---  &  8.0  &  ---  &  ---  &  0.039 \\
SDSSJ090202.19+101759.7  & $ -0.14 $ &  8.34  &  ---  &  ---  &  5.0  &  ---  &  ---  &  0.048 \\
UGC00272  & $ -0.08 $ &  8.40  &  ---  &  ---  &  7.4  &  ---  &  ---  &  0.049 \\
UGC01333  & $ -0.12 $ &  8.69  &  ---  &  ---  &  28.4  &  ---  &  ---  &  0.052 \\
UGC01395  & $ -0.07 $ &  8.63  &  ---  &  ---  &  10.1  &  ---  &  ---  &  0.059 \\
UGC02019  & $ -0.08 $ &  8.61  &  0.4  &  ---  &  7.6  &  0.2  &  ---  &  0.030 \\
UGC03634$\;^{\dagger\,\star}$ & $ -0.01 $ &  8.62  &  ---  &  ---  &  ---  &  ---  &  ---  &  --- \\
UGC05378  & $ -0.13 $ &  8.67  &  0.7  &  ---  &  4.1  &  0.5  &  ---  &  0.047 \\
UGC05691  & $ -0.18 $ &  8.73  &  0.8  &  ---  &  26.5  &  0.6  &  ---  &  0.051 \\
UGC06332$\;^{\dagger\,\star}$ & $ 0.00 $ &  8.59  &  ---  &  ---  &  ---  &  ---  &  ---  &  --- \\
UGC09530  & $ -0.31 $ &  8.70  &  0.3  &  ---  &  10.8  &  0.4  &  ---  &  0.083 \\
UGC11001  & $ -0.18 $ &  8.61  &  0.3  &  1.2  &  5.7  &  0.4  &  1.4  &  0.027 \\
UGC11214  & $ -0.11 $ &  8.56  &  ---  &  ---  &  2.6  &  ---  &  ---  &  0.070 \\
UGC11816  & $ -0.26 $ &  8.77  &  ---  &  ---  &  5.2  &  ---  &  ---  &  0.083 \\
UGC12158  & $ -0.09 $ &  8.65  &  ---  &  ---  &  10.6  &  ---  &  ---  &  0.065 \\
\label{table2}
\end{longtable}
\end{longtab}
\end{center}
\newpage

\section{On the use of different calibrators in the derivation of the abundance profiles}\label{sec:appendix3} 

In this appendix we assess the effect of using different calibrators on the results presented in the paper. With such purpose, we replicate the figures that support our main results using the calibration described in \citet[][D16]{dopita2016}, and the one proposed in \citet[][]{marino2013} for the N2 index (M13-N2, see Sect.~\ref{sec:analysis2c} for a brief description of these two additional calibrators). The choice of these empirical indicators is mainly based on the limited wavelength range in the blue regime covered by MUSE, that does not include more widely used emission lines such as [\oii]$\lambda3727$.

\begin{figure}[h]
\begin{center}
\resizebox{0.69\hsize}{!}{\includegraphics{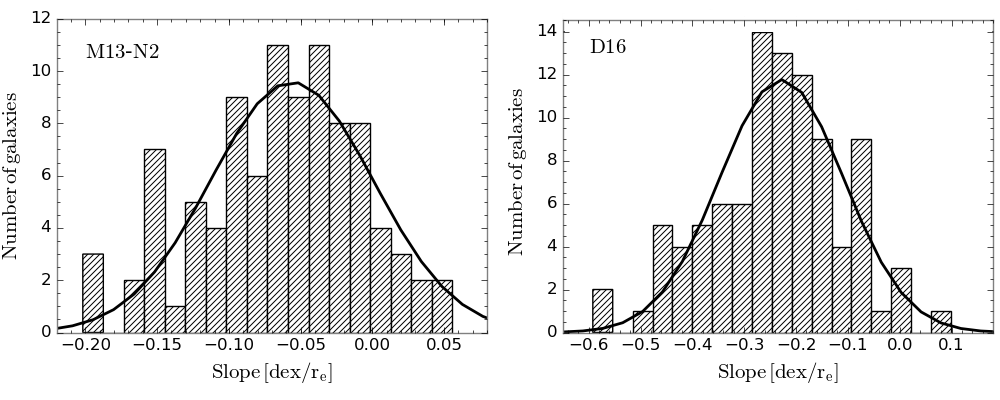}}
\caption{Distribution of slope values of the oxygen main abundance gradients (normalised to the disc effective radius) derived for the galaxy sample using the M13-N2 ({\it left}) and D16 ({\it right}) calibrators. Solid black line represents the Gaussian distribution of the data assuming the mean and standard-deviation of the distribution of slope values and sampled with the same bins.}
\label{fig:slopes_others}
\end{center}
\end{figure}

Figure~\ref{fig:slopes_others} shows the distribution of slopes of the main oxygen abundance gradients derived for the galaxy sample of 95 spiral galaxies using the M13-N2 ({\it left}) and D16 ({\it right}) calibrators. As obtained with O3N2, the abundance distribution in both cases presents a characteristic slope very similar for all galaxies, although the actual value changes ($\rm -0.23 \,dex/r_e$ for D16 and $\rm -0.06 \,dex/r_e$ for M13-N2). 

\begin{figure}[h]
\centering
\resizebox{0.9\hsize}{!}{\includegraphics{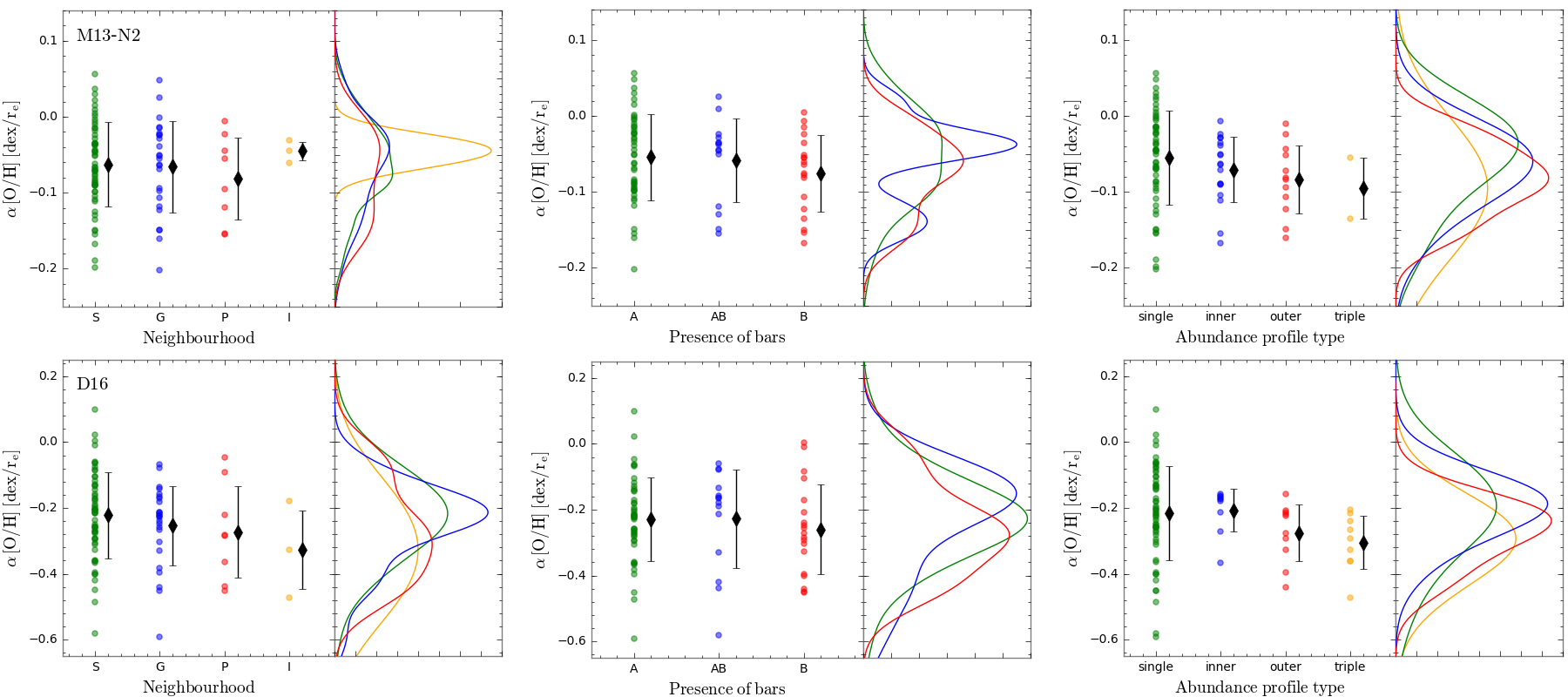}}
\caption{Distribution of slopes depending on different galaxy properties using the M13-N2 ({\it top panels}) and the D16 ({\it bottom panels}) calibrators: (i) the environment of the galaxies (isolated, green; in groups, blue; paired, red; interacting, yellow) for the {\it left panel}, (ii) the presence or absence of bars (clearly unbarred galaxies, green; clearly barred, red; and an intermediate stage, blue) for the {\it middle panel}, and (iii) the shape of the abundance profile (single profile, green; broken profile with inner drop, blue; broken profile with outer flattening, red; and a doubly-broken profile with presence of both inner drop and outer flattening, yellow) for the {\it right panel}. The black diamonds represent the mean values within each segregation, with the error bars indicating the standard deviation.}
\label{fig:slope_dependences_others}
\end{figure}

We have also analysed the influence of some galaxy properties such as the density of the environment, the presence of bars and the type of abundance profile displayed (according to Sect.~\ref{sec:shape}) on the main abundance gradient when we use these two calibrators. The outcome of this analysis is shown in Fig.~\ref{fig:slope_dependences_others}. Similarly to O3N2, we only find a dependence of the slope with the shape of the abundance profile, presenting the single profiles the shallowest gradients and the doubly-broken profiles the steepest ones. The mean and standard deviation values of the distributions for the M13-N2 (D16) are the following ones (the nomenclature is explained in Sect~\ref{sec:slope}):
\begin{gather*}
\hspace{0.1cm}{\rm S:}\hspace{8pt} \alpha_{O/H} = -0.06\,(-0.22) \,\rm{dex}/r_e,\,\, \sigma = 0.06\,(0.13) \,\rm{dex}/r_e \hspace{50pt} {\rm A:}\hspace{13pt} \alpha_{O/H} = -0.05\,(-0.23) \,\rm{dex}/r_e,\,\, \sigma = 0.06\,(0.13) \,\rm{dex}/r_e \\\relax
\hspace{0.1cm}{\rm G:}\hspace{6pt} \alpha_{O/H} = -0.07\,(-0.25) \,\rm{dex}/r_e,\,\, \sigma = 0.06\,(0.12) \,\rm{dex}/r_e \hspace{50pt} {\rm AB:}\hspace{6pt} \alpha_{O/H} = -0.06\,(-0.22) \,\rm{dex}/r_e,\,\, \sigma = 0.06\,(0.15) \,\rm{dex}/r_e\\\relax
\hspace{0.1cm}{\rm P:}\hspace{7pt} \alpha_{O/H} = -0.08\,(-0.27) \,\rm{dex}/r_e,\,\, \sigma = 0.05\,(0.14) \,\rm{dex}/r_e \hspace{51pt} {\rm B:}\hspace{13pt} \alpha_{O/H} = -0.08\,(-0.26) \,\rm{dex}/r_e,\,\, \sigma = 0.05\,(0.14) \,\rm{dex}/r_e\\\relax
\hspace{0.1cm}{\rm I:}\hspace{10pt} \alpha_{O/H} = -0.05\,(-0.33) \,\rm{dex}/r_e,\,\, \sigma = 0.01\,(0.12) \,\rm{dex}/r_e.
\end{gather*}
\begin{gather*}
\hspace{2.8cm}{\rm Single:}\hspace{48.5pt} \alpha_{O/H} = -0.06\,(-0.22) \,\rm{dex}/r_e,\, \sigma = 0.06\,(0.14) \,\rm{dex}/r_e \,; \hspace{5pt}n_{gal} = 55\,(59)\\\relax
\hspace{2.8cm}{\rm Broken \,(inner):}\hspace{15.5pt} \alpha_{O/H} = -0.07\,(-0.21) \,\rm{dex}/r_e,\, \sigma = 0.04\,(0.07) \,\rm{dex}/r_e \,; \hspace{5pt}n_{gal} = 18\,(9)\\\relax 
\hspace{2.8cm}{\rm Broken \,(outer):}\hspace{15.5pt} \alpha_{O/H} = -0.08\,(-0.28) \,\rm{dex}/r_e,\, \sigma = 0.05\,(0.09) \,\rm{dex}/r_e \,; \hspace{5pt}n_{gal} = 12\,(10)\\\relax 
\hspace{2.8cm}{\rm Doubly-broken:}\;\; \alpha_{O/H} = -0.10\,(-0.31) \,\rm{dex}/r_e,\, \sigma = 0.04\,(0.08) \,\rm{dex}/r_e \,; \hspace{5pt}n_{gal} = 2\,(9).
\end{gather*}

\begin{figure}[h]
\centering
\resizebox{0.69\hsize}{!}{\includegraphics{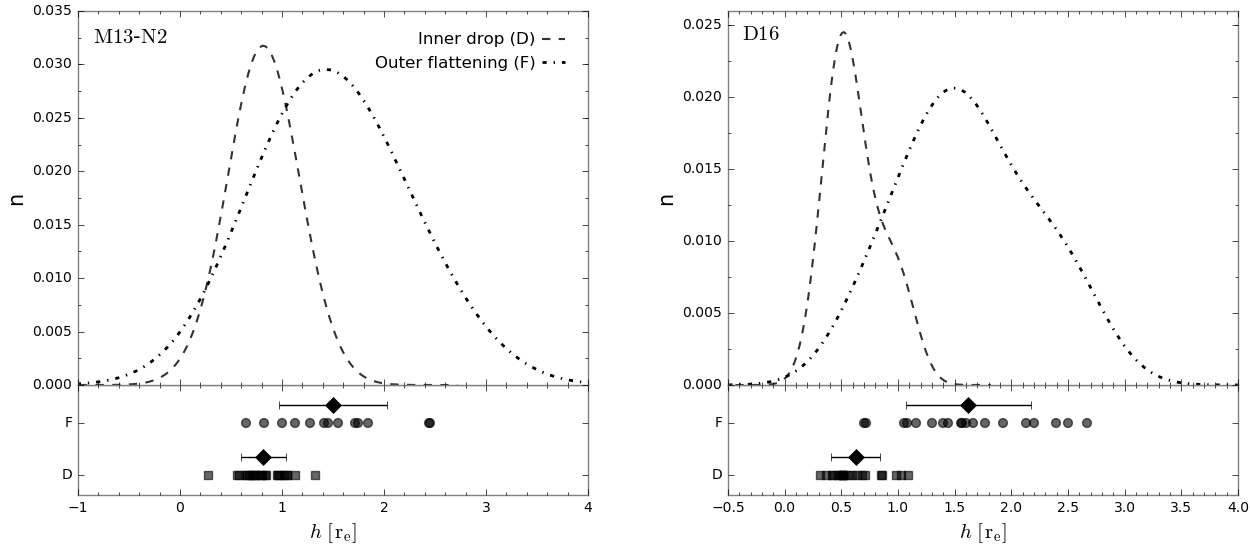}}
\caption{Distribution of radial positions (in units of $\rm r_e$) of the inner drop (D, squares) and outer flattening (F, dots) found in some galaxies of the sample using the M13-N2 ({\it left panel}) and the D16 ({\it right panel}) calibrators. The black diamonds represent the mean values of the position at which these features appear, with the error bars indicating the standard deviations. Dashed and dashed-dotted lines represent the normalised density distributions of the location of the inner drop and outer flattening, respectively.}
\label{fig:dropflat_others}
\end{figure}

Regarding the presence of inner drops and outer flattenings in the abundance profiles, the use of the new calibrators reduces the number of galaxies where the inner abundance drop is detected to 23 for M13-N2 and to 20 for D16. The number of galaxies displaying the outer abundance flattening is also reduced to 13 galaxies in the case of the M13-N2 calibrator and to 19 in the case of D16. Analysing the distribution of radial positions of both features we obtain a very similar shape to the one obtained with O3N2 (see Fig.~\ref{fig:dropflat_others}): a narrow distribution for the drop and a quite wider one for the flattening. The average positions of the features are also very similar to the previous case, with only small differences: the inner drop is located at $\rm \sim 0.8 \,r_e$ (standard deviation of 0.22) for M13-N2 and at $\rm \sim 0.6 \,r_e$ (standard deviation of 0.22) for D16, and the average position of the outer flattening is $\rm \sim 1.5 \,r_e$ (standard deviation of 0.53) and $\rm \sim 1.6 \,r_e$ (standard deviation of 0.55), respectively.

\begin{figure}[!h]
\centering
\resizebox{0.75\hsize}{!}{\includegraphics{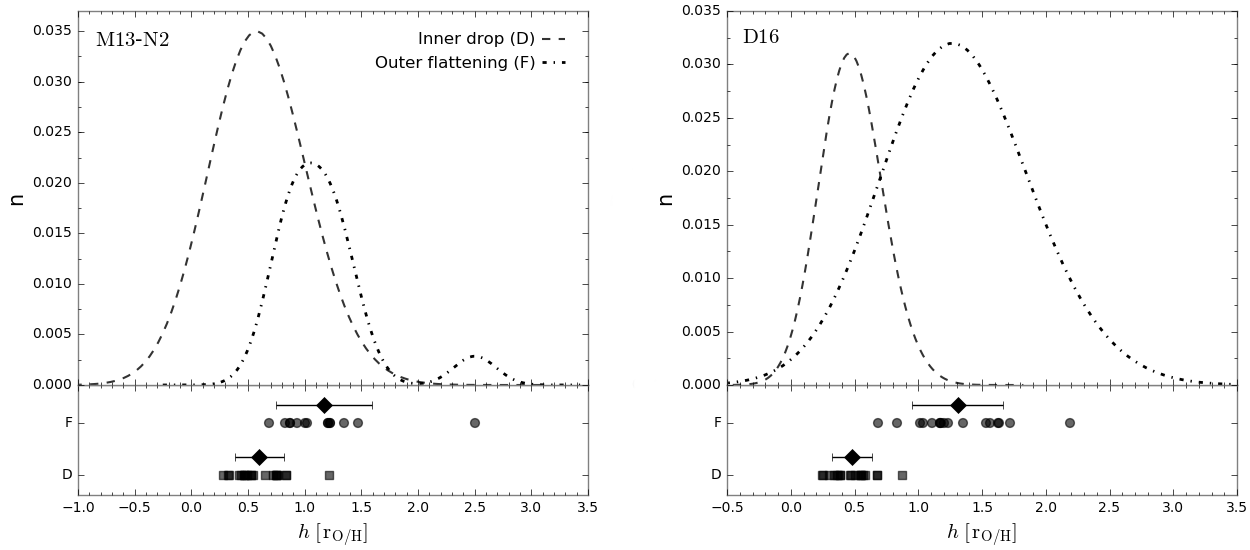}}
\caption{Same as Fig.~\ref{fig:dropflat_others} but normalising the abundance distribution to r$_{O/H}$. See caption above for more details.}
\label{fig:dropflatralpha_others}
\end{figure}

\begin{figure}[!h]
\centering
\resizebox{0.8\hsize}{!}{\includegraphics{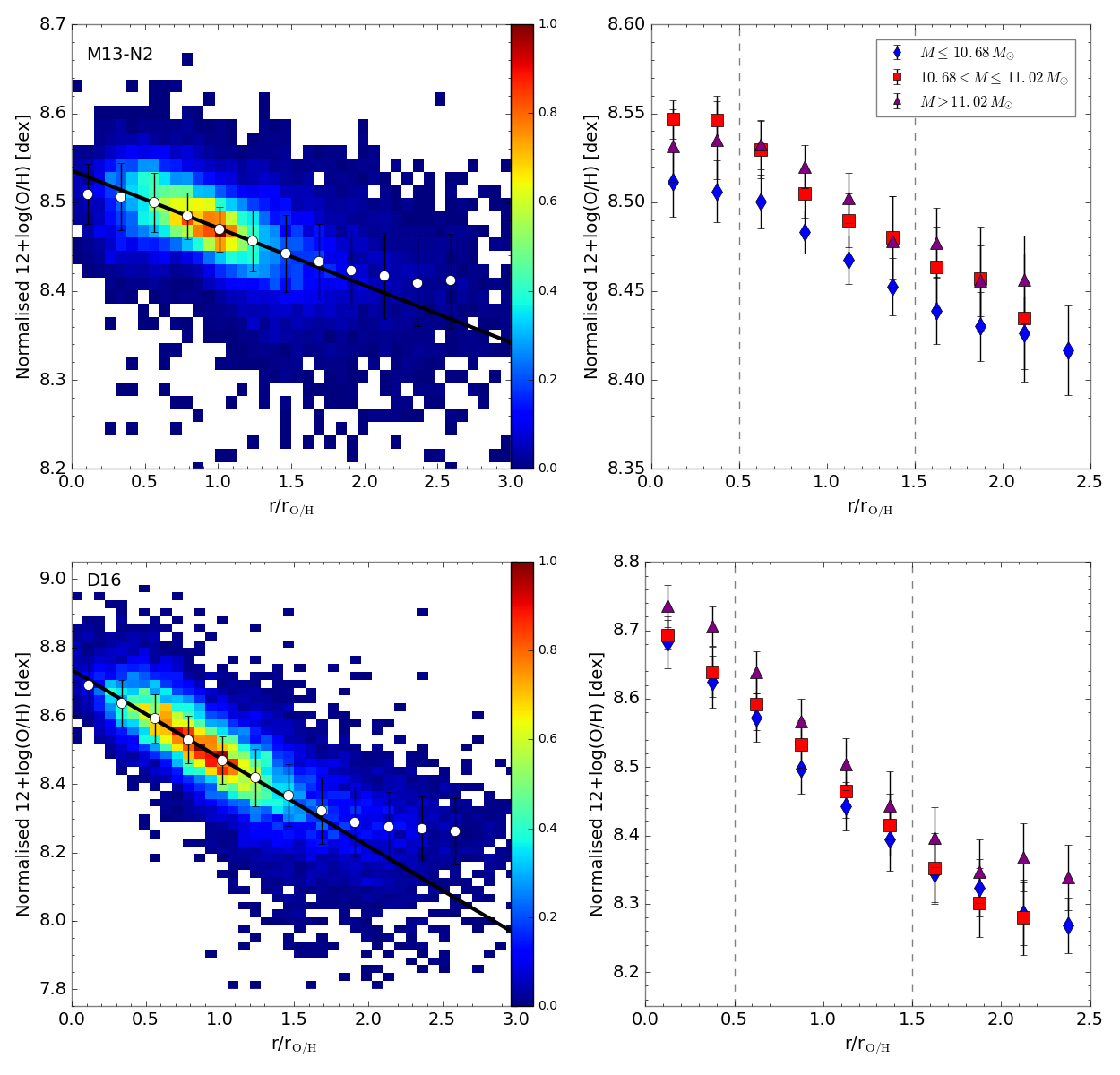}}
\caption{{\it Left}: Normalised radial density distribution of the oxygen abundance for the M13-N2 ({\it top}) and the D16 ({\it top}) calibrators. The galactocentric distances are normalised to the abundance scalelength r$_{O/H}$. The white dots represent the mean oxygen abundance values, with the error bars indicating the corresponding standard deviations, for bins of 0.3 r$_{\rm e}$. The solid black line represents the error-weighted linear fit derived for those mean values within the range between 0.5 and 1.5 r$_{\rm e}$. {\it Right}: Mean oxygen abundance radial profiles derived for galaxies belonging to three different stellar mass bins: $\log (M/M_\odot) \leq 10.69$, blue diamonds; $10.69 < \log (M/M_\odot) \leq 11.05$, red squares; $\log (M/M_\odot) > 11.05$, purple triangles. The limits of the bins were chosen to ensure a similar number of elements in each bin.  The symbols represent the mean oxygen abundance values, with the error bars indicating the corresponding standard deviations, for bins of 0.25 r$_{O/H}$. Dashed vertical lines indicate the average position of the inner drop and the flattening in the outer parts.}
\label{fig:commongrad_ralpha_others}
\end{figure}

When normalising the abundance profiles to the abundance scale length $r_{\rm O/H}$ (see Sect.~\ref{masses} for information about this parameter), the distributions of radial positions for these features get significantly narrower (standard deviations are reduced from 0.22 to 0.17 and from 0.22 to 0.16 for the inner drop, from 0.53 to 0.22 and from 0.55 to 0.35 for the outer flattening, using the M13-N2 and the D16 calibrators, respectively), deriving very similar values for the position of the inner drop and the outer flattening (see Fig.~\ref{fig:dropflatralpha_others}). In addition, we reproduce the stacking of the abundance distribution with this new normalisation scale in Fig.~\ref{fig:commongrad_ralpha_others}. We find a very similar shape for the two calibrators. By construction, the global abundance distribution presents a characteristic slope of $-0.20 \,r_{\rm O/H}$ in the case of the M13-N2 calibrator and $-0.04 \,r_{\rm O/H}$ for D16. In addition, the abundance distribution shows clear evidences of the presence of the inner drop and the outer flattening, as in the case of the M13-O3N2 calibrator. The right panel of Fig.~\ref{fig:commongrad_ralpha_others} shows the average abundance radial profiles when separating the sample into different mass bins for both the M13-N2 and D16 calibrators. It can be seen that the presence of the abundance drop in the inner regions of the galaxy discs and the flattening in the outer regions is also easily observable, although for D16 the presence of the inner drop is not so clear. Similarly to M13-O3N2, we find a dependence of the presence of the inner drop with the galaxy mass, with the most massive galaxies presenting the deepest abundance drops. The flattening on the opposite appear independently of the galaxy mass.

In conclusion, in this appendix we demonstrate that the main results presented in this paper do not depend on the choice of the oxygen abundance indicator used. However, we should be aware of the subtle differences that appear when using different calibrators, especially if we focus on the abundance distribution of individual galaxies.

\clearpage

\section{Graphic characterisation of the sample}\label{sec:appendix4} 

In this appendix we include a graphic characterisation of the galaxies in the sample. Figure~\ref{fig:characterisation} shows two RGB colour images of the galaxies, using SDSS $r$-, $i$- and $z$-band images recovered from the data (in red, green and blue respectively, {\it left}) and narrow-band images centred in the emission lines \mbox{[\nii]~$\lambda6584$} in red, H$\alpha$ in green, and \mbox{[\oiii]~$\lambda5007$} in blue ({\it middle-left}). The {\it middle-right} panels present the H$\alpha$ map of the galaxies together with the ionised regions detected by {\scshape HIIexplorer} (with $\rm S/N>3$) shown as black segmented contours. Finally, the {\it right} panels show the deprojected oxygen abundance radial distribution of the \hii\,regions (violet circles) normalised to the disc effective radius. The solid blue line represents the fit to this distribution, and the dashed vertical lines correspond to the radial position of the inner drop and/or outer flattening, if present. The slope ($a$) and zero-point ($b$) of the main negative abundance gradient are indicated in the top-right corner. 

\vspace{0.5cm}
\noindent Abridged version of the appendix. Full version available in the A\&A webpage or by request (lsanchez@iaa.es).

\begin{figure}[!h]
\centering
\caption{Characterisation of the galaxies in the sample. See text above for details.}
\resizebox{\hsize}{!}{\includegraphics{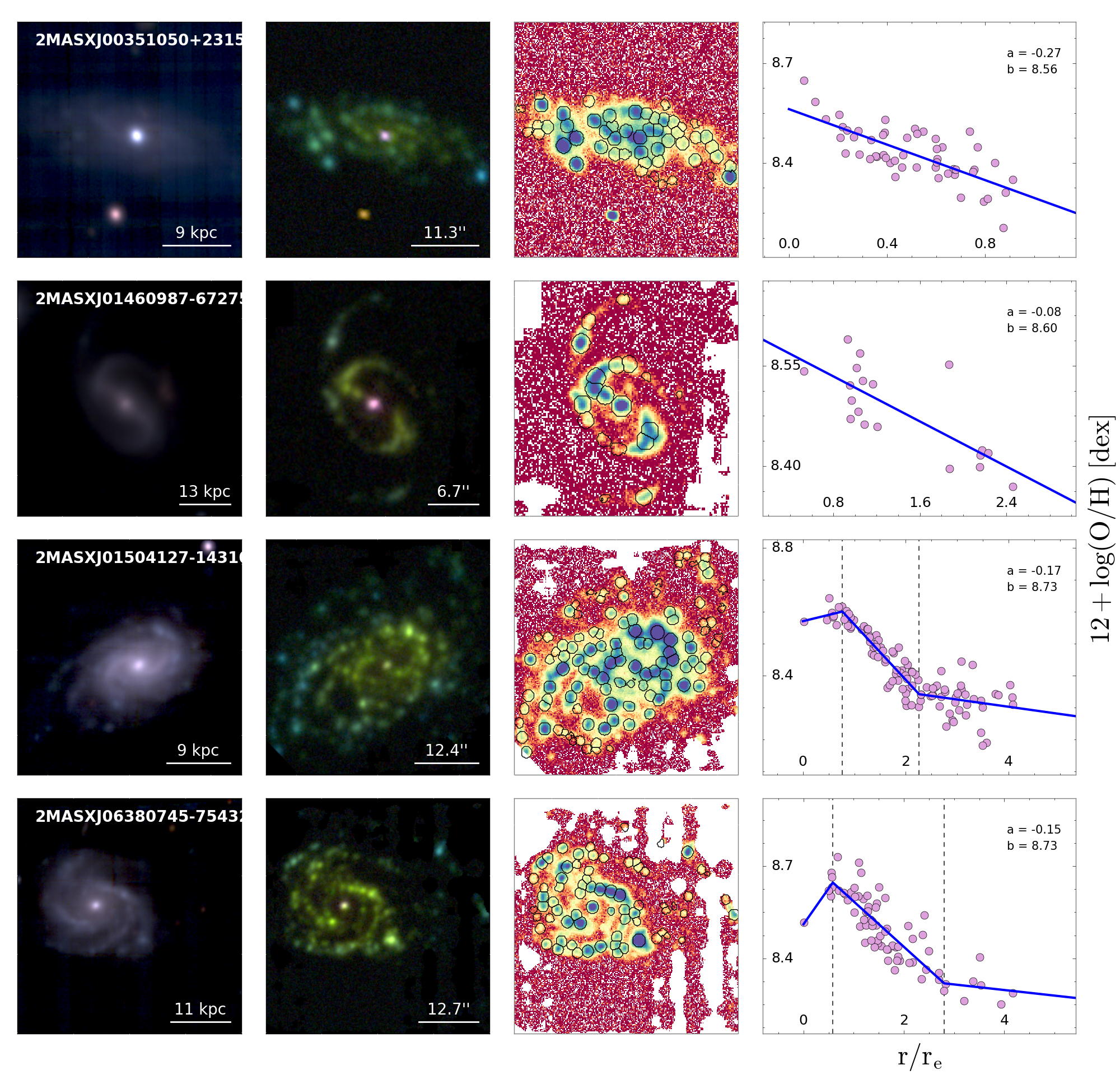}}
\label{fig:characterisation}
\end{figure}





\setcounter{figure}{0}
\newpage

\begin{figure}
\centering
\caption{Continued. Characterisation of the galaxies in the sample. See text above for details.}
\resizebox{\hsize}{!}{\includegraphics{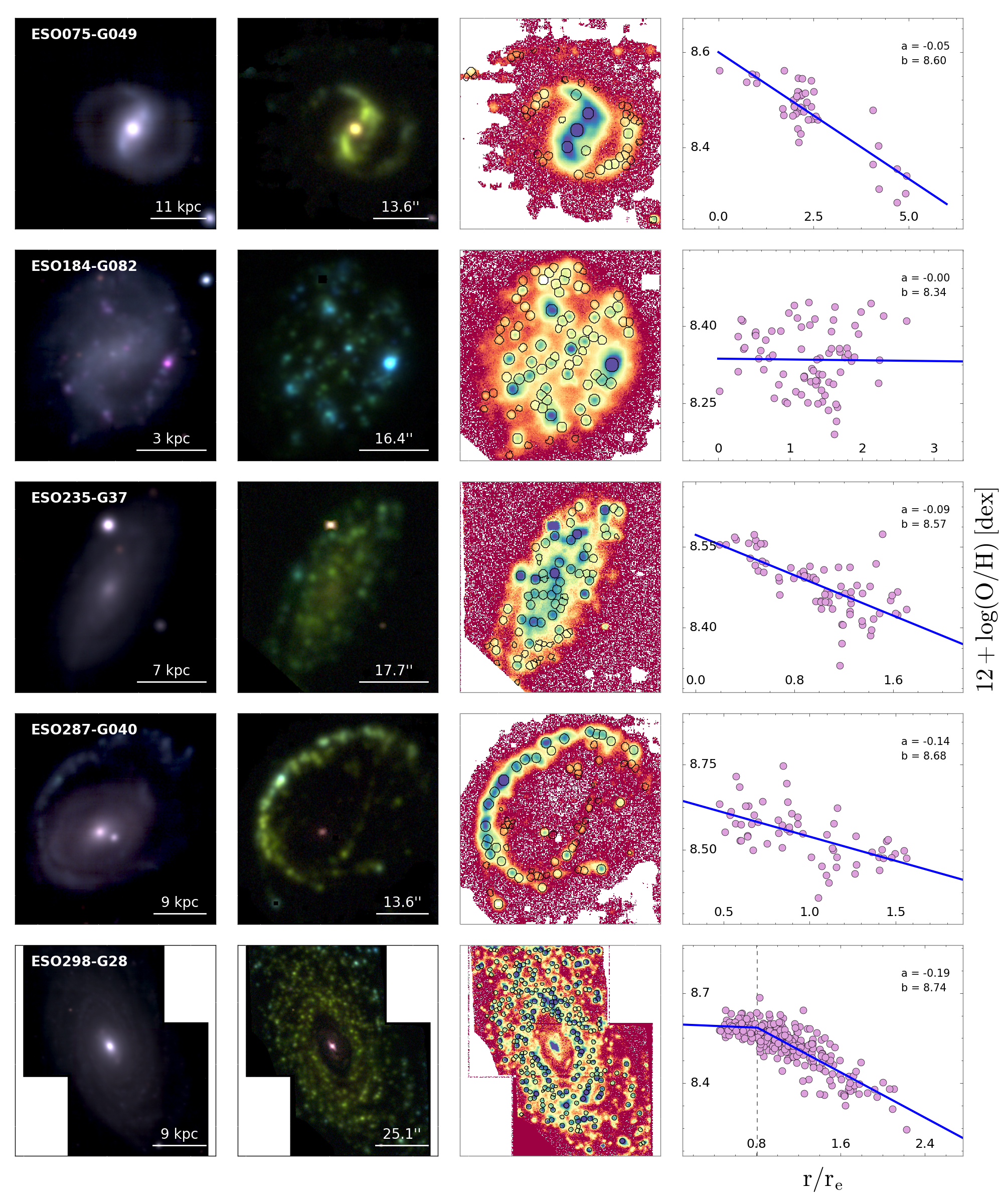}}
\label{fig:characterisation}
\end{figure}

\setcounter{figure}{0}
\newpage

\begin{figure}
\centering
\caption{Continued. Characterisation of the galaxies in the sample. See text above for details.}
\resizebox{\hsize}{!}{\includegraphics{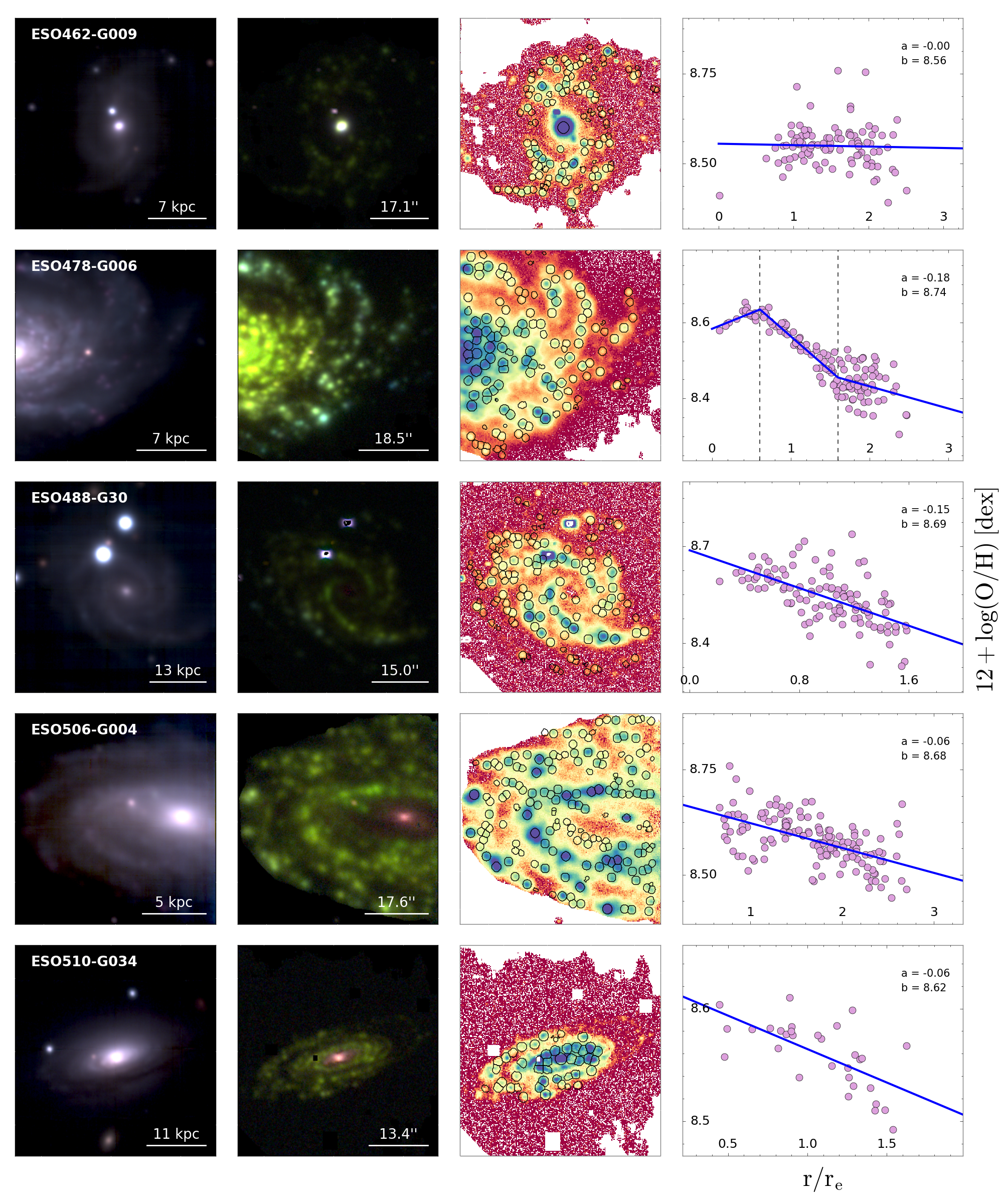}}
\label{fig:characterisation}
\end{figure}

\end{document}